\documentclass[nofootinbib,aps,prb,a4paper,twocolumn,english,superscriptaddress,longbibliography,reprint]{revtex4-2}
\usepackage{verbatim}
\usepackage{color}
\usepackage{tikz}
\usepackage{bbm}
\usetikzlibrary{positioning, arrows.meta}

\usepackage{graphicx}% Include figure files
\usepackage{dcolumn}% Align table columns on decimal point
\usepackage{bm}
\usepackage{appendix}
\usepackage{mathtools}
\usepackage{amsmath,amssymb,amsfonts,mathrsfs, amsbsy}
\usepackage[colorlinks=true, linkcolor=blue, citecolor=blue]{hyperref}
\usepackage{soul}

\newcommand{\ubm}[1]{\bm{\mathrm{#1}}} % Defines the command \ubm
\renewcommand{\vec}[1]{\boldsymbol{\mathbf{#1}}}

\newcommand{\norm}[1]{\lVert #1 \rVert}

\begin{document}

\preprint{APS/123-QED}

\title{Anomalous transport in non-integrable classical field theories}% Force line breaks with \\

\author{Matija Koterle}
\email{matija.koterle@fmf.uni-lj.si}
\affiliation{%
 Department of Physics, Faculty of Mathematics and Physics,
University of Ljubljana, Jadranska 19, SI-1000 Ljubljana, Slovenia%\\
}%
\author{Toma\v z Prosen}%
\affiliation{%
 Department of Physics, Faculty of Mathematics and Physics,
University of Ljubljana, Jadranska 19, SI-1000 Ljubljana, Slovenia%\\
}%
\author{Tianci Zhou}%
\email{tzhou13@vt.edu}
\affiliation{Department of Physics, Virginia Tech, Blacksburg, VA, USA 24061}
\affiliation{Virginia Tech Center for Quantum Information Science and Engineering, Blacksburg, VA 24061, USA}

\date{\today}% It is always \today, today,

\begin{abstract}
Anomalous KPZ spin transport is well established in integrable non-Abelian lattice models but has not been investigated in continuum field theories as discretization in numerics generally break the continuum theory's integrability. We show that finite temperature acts as a regulator that can restore anomalous transport over a broad time window. In a family of spin field theories labeled by integer $n$, the $n = 1$ case is the Landau-Lifshitz model, whose numerical data shows spin superdiffusion with
Kardar-Parisi-Zhang (KPZ) scaling and, at lower temperature ballistic energy transport, whereas both observables are diffusive at high temperature. The non-integrable $n = 2$ case shows the same crossover. While Lyapunov analysis confirms the model's non-integrability, the structure of spin-density space-time profiles suggests that long-lived soliton-like trajectories exist at low temperature. 
\end{abstract}

\maketitle

\newpage

%\section{\label{sec:intro}Introduction\\}
\emph{Introduction.---} The anomalous superdiffusive spin transport in integrable spin chains with non-abelian symmetries has attracted significant theoretical and experimental 
interest~\cite{vznidarivc2011transport,prosen2013macroscopic,ljubotina_2019_kardar,
krajnik2020kardar,krajnik2020integrable,
dupont_2020_universal,scheie_2021_detection,
ilievski2021superuniversality,bulchandani_2021_superdiffusion,bertini_2021_finite,
wei_2022_quantum,rosenberg_2024_dynamics,keenan_2023_evidence,krajnik_2024_dynamical,de_nardis_anomalous_2019}. Prototypical models, such as the lattice Landau-Lifshitz spin chain ~\cite{sklyanin1979complete, ishimori1982integrable, haldane1982excitation}, exhibit a superdiffusive length scale that grows as $t^{\frac{2}{3}}$ and the correlation of spin collapses to the Pr\"ahofer-Spohn scaling function~\cite{prahofer2004exact}. The phenomenon is ``super-universal'' in integrable models with a non-abelian symmetry~\cite{krajnik2020integrable,ilievski2021superuniversality}, appearing both quantum~\cite{vznidarivc2011transport,ljubotina_2019_kardar,dupont2020universal} and classical systems~\cite{prosen2013macroscopic,krajnik2020kardar,takeuchi2025partial}, as well as for spin chains with higher-rank non-abelian groups~\cite{ye2022universal}. Existing studies overwhelmingly focus on discrete lattice models. In this work, we investigate whether such spin superdiffusive phenomenology persists in the corresponding (classical) integrable non-abelian continuum field theories.

Na\"ive discretizations of integrable continuum models typically break integrability, and thus a direct numerical investigation may fail to reproduce the underlying integrable dynamics. For instance, a simple lattice formulation of the Landau-Lifshitz (LL) field theory yields the non-integrable classical Heisenberg chain ($H = \sum_i \vec{m}_i \cdot \vec{m}_{i+1}$ where $\vec{m}_i$ are classical unit length spin vectors)); restoring its integrability requires a specific logarithmic deformation of interaction \cite{sklyanin1979complete, ishimori1982integrable, haldane1982excitation}. The effects of integrability breaking on the lattice are negligible in the dynamics of spin configurations which are sufficiently smooth. However, when the canonical ensemble average is introduced to compute transport phenomena, the thermally fluctuating spin configurations on the lattice are presumed to break the integrability.

We propose to use temperature itself as a physical regulator to mitigate integrability breaking. A thermal state inherits a thermal length scale $1/\beta$, below which fluctuations of the configuration are suppressed. Thus, it opens a temperature dependent time window that allows us to observe superdiffusive dynamics if it is indeed an intrinsic feature of the continuum theory.
We apply this thermal regulator to a family of continuum models governed by the Hamiltonian
\begin{equation}\label{eq:hamiltonian}
H^{(n)} = \frac{1}{2n} \int_{-L/2}^{L/2} ( \partial_x \vec{m} \cdot \partial_x \vec{m})^n \mathrm{d}x 
\end{equation}
where $\vec{m}\equiv \vec{m}(x)$ is a classical unit vector spin field on the real axis and $n$ is a positive integer. The $n = 1$ case corresponds to the integrable Landau-Lifshitz (LL) field theory at the isotropic point \cite{sklyanin1979complete,takhtajan_1977_integration,faddeev_1987_hamiltonian}. 
At high temperature (e.g. $\beta = 0$) and for standard discretization (i.e. UV cutoff) as discribed below, the dynamics resembles that of a generic non-integrable model, exhibiting diffusion of magnetization and energy. In contrast, in a long time window at low temperature ($\beta\sim1$), we recover spin superdiffusion with a dynamical exponent $\frac{3}{2}$ and correlation function described by the KPZ scaling function \cite{ljubotina_2019_kardar,takeuchi2025partial}. The ballistic transport of energy is recovered only at a much lower temperature ($\beta>10$).

This crossover is consistent with the known effects of integrability-breaking perturbations in lattice models~\cite{ferreira2020ballistic, de2021stability, roy2023robustness, mccarthy2024slow, mcroberts2024parametrically, wang2025breakdown}. In such systems, a small integrability breaking perturbation with strength $g$ induces a crossover time scale of $t^* \sim g^{-6}$ \cite{mccarthy2024slow, mcroberts2024parametrically,wang2025breakdown,roy_robustness_2023} below which integrable dynamics reemerge. Our discretized field theory at finite temperature can be viewed as the continuum limit with an added symmetric integrability breaking perturbation from discretization which is effectively weak at low temperature. The thermal regulator cures the integrability breaking within the timescale $t^*$. 

\begin{figure*}[t]
  \centering
  \includegraphics[width=\textwidth]{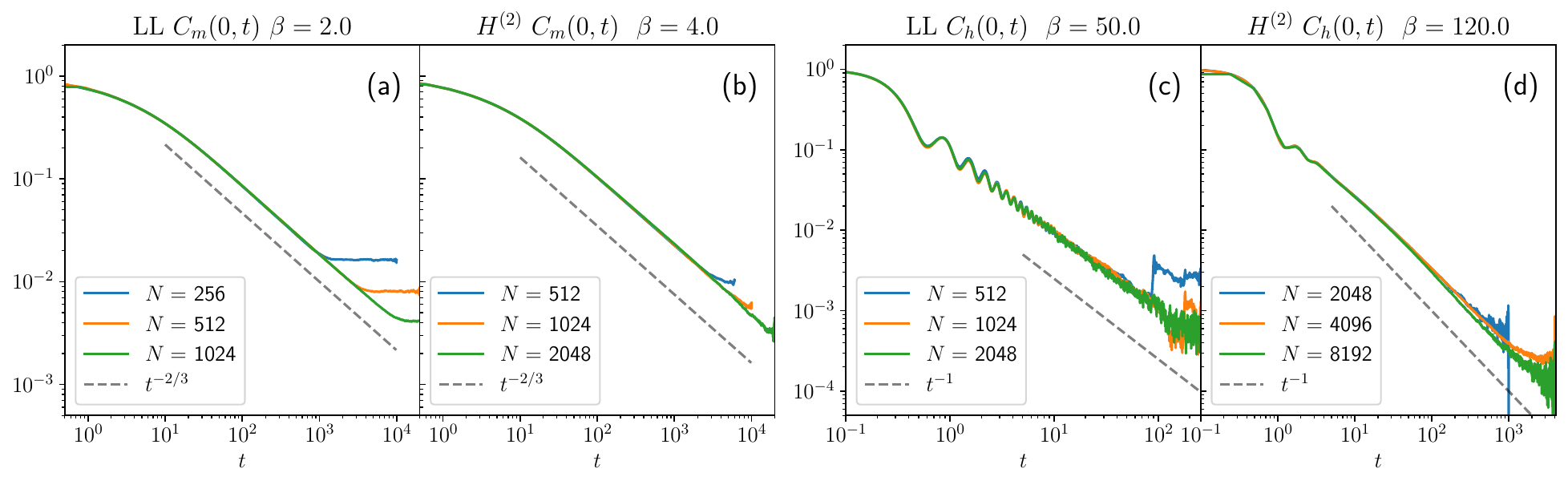}
  \caption{\textbf{(a-b)} Transport of magnetization in the LL and non-integrable $n=2$ model at finite $\beta$. Both cases exhibit superdiffusive transport with a dynamical exponent $z_m=3/2$.
  \textbf{(c-d)} Transport of energy in the LL and non-integrable $n=2$ models at finite $\beta$. While both cases exhibit ballistic transport, it is qualitatively different. LL shows large oscillations with revivals at times twice of system size (finite size effect), while the non-integrable model slowly approaches $z_h=1$ only at timescales $t>100$.}
  \label{fig:mm_hh_t}
\end{figure*}

The somewhat surprising discovery arises for $n = 2$. The $n = 2$ model is not known to be close to any integrable point, yet it exhibits a similar crossover in spin and energy transport. At high temperature, the magnetization exhibits the expected diffusive behavior, while the transport of energy exhibits a diffusive central peak with exponentially suppressed ballistic tails (see End Matter). At lower temperatures, we again observe the superdiffusive magnetization transport, and at an even lower temperature, ballistic transport of energy. Do these observations suggest that the $n = 2$ model at low temperature effectively approaches a hidden integrable point, or is anomalous transport in field theories more generic than in lattice models?

Although a complete answer is for further investigation, we numerically probe the integrability of the $n = 2$ model. We measure the finite-time Lyapunov exponents (FTLE) in a thermal ensemble. For both the $n = 1$ and $n = 2$ models, the distance between initially close trajectories grows exponentially. However, the $n = 1$ model exhibits a much smaller FTLE at lower temperatures, which is consistent with its near-integrable feature. We further produce the heatmaps of the third component of magnetization density $m^3(x,t)$ for a single trajectory. At low temperature, we clearly observe long-lived soliton-like trajectories in both models, which echoes the findings in discrete non-integrable spin chains~\cite{mcroberts2022anomalous,mcroberts2022long,mcroberts2024parametrically}. Despite its non-integrability, these slowly moving quasi-particles in the $n = 2$ model can be responsible for the superdiffusive spin transport. The ballistic energy transport in such systems can be understood as a consequence of momentum conservation in the continuum, where the overlap of the Hamiltonian and the momentum current leads to a non-zero Drude weight.

% \section{\label{sec:models}Models}
% \noindent 
{\it Models.---} We study a class of one-dimensional classical field theories on a domain $x \in [-L/2, L/2)$ with periodic boundary conditions. The fundamental degree of freedom is a three-component classical spin field $\ubm{m}(x,t) = (m^1, m^2, m^3)$ constrained to have unit length, $|\ubm{m}(x,t)|^2 = 1$. The dynamics are governed by the canonical Poisson bracket relations:
\begin{equation}
    \{m^\alpha(x), m^\beta(y) \} = \varepsilon^{\alpha \beta \gamma} m^\gamma(x) \delta(x-y),
\end{equation}
where $\varepsilon^{\alpha\beta\gamma}$ is the Levi-Civita symbol. We consider a class of Hamiltonians indexed by a positive integer $n$ as defined in Eq.~\eqref{eq:hamiltonian}. The equation of motion for the spin-field dynamics is of the Landau-Lifschitz type
\begin{equation}\label{eq:dynamics}
    \frac{\partial\ubm{m}}{\partial t} = \ubm{m} \times \frac{\delta H^{(n)}}{\delta \ubm{m}} = - \, \ubm{m} \times \partial_x \left[ \left( \partial_x \ubm{m} \cdot \partial_x \ubm{m} \right)^{n-1} \partial_x \ubm{m} \right].
\end{equation}
We study the transport properties of the system by numerically investigating the transport of densities of conserved quantities. A conserved quantity, $Q = \int q(x) \, \mathrm{d}x$, is the spatial integral of a local density $q(x,t)$. We compute the connected two-point autocorrelation function of the density field
\begin{equation}
    C_q(x, t) = \langle q(x, t) q(0, 0) \rangle_\beta^c,
\end{equation}
where $\langle \cdot \rangle_\beta^c$ denotes a connected thermal average over the Gibbs ensemble at inverse temperature $\beta$. The conservation of $Q$ dictates that this integral $\int C_q(x,t) \, \mathrm{d}x$ remains constant in time. 

\begin{figure*}[t!]
  \centerline{
  \includegraphics[width=1.0\textwidth]{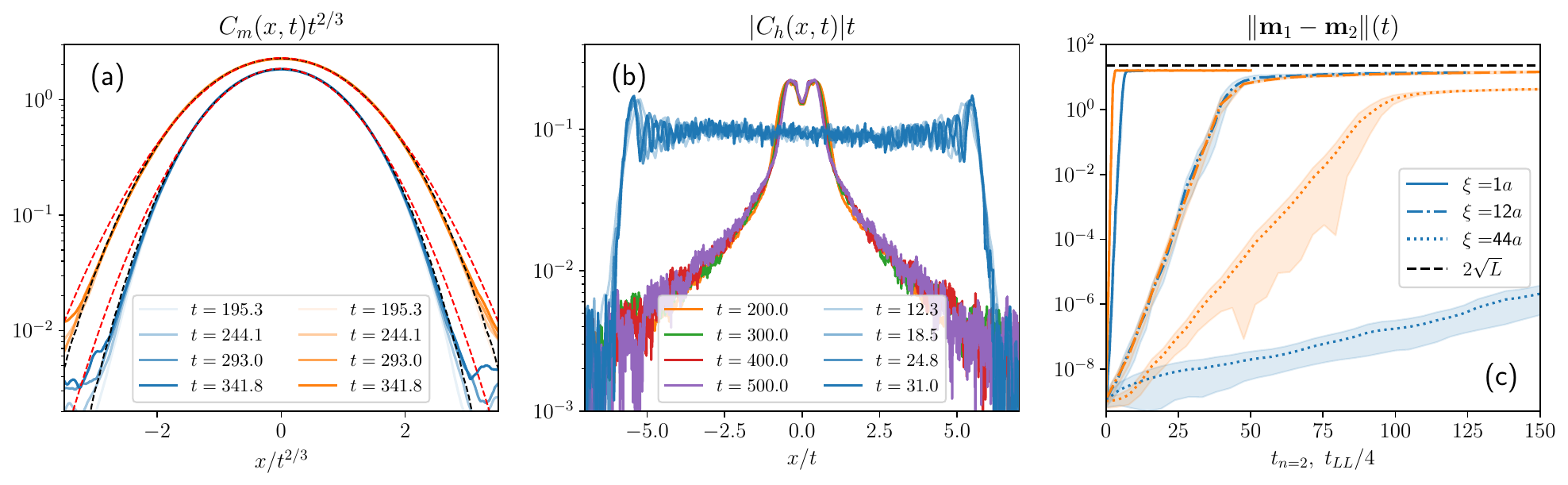}
  }
  \caption{Normalized scaling functions in rescaled coordinates. The blue curves show dynamics generated by LL, while the orange shows dynamics of the $n=2$ model. \textbf{(a)} Magnetization transport (LL at $\beta=2$, $N=1024$, $n=2$ at $\beta=4$, $N=2048$). The dashed lines denote fits of a Gaussian and Pr\"ahofer-Spohn (red and black dashes respectively) scaling functions fitted to $C_m(x,t)$ at the latest displayed times. \textbf{(b)} Energy transport (LL at $\beta=50$ with $N=2048$, $n=2$ at $\beta=120$ with $N=16384$). The colored lines show results for $n=2$ model to showcase the data collapse with a ballistic rescaling. \textbf{(c)} Comparison of exponential deviations of trajectories in thermal ensembles for the LL ($\beta\in\{0,3,11\}$, at times corresponding to top axis) and $n=2$ ($\beta\in\{0,4,40\}$ at times corresponding to bottom axis). The choices of $\beta$ fix equal correlation lengths in both models of $\{1,12,44\}$ lattice spacings $a$ respectively. The norm is upper bounded by $2\sqrt{L}$. Integrator tolerance $10^{-10}$. Note that the $x$ coordinates are rescaled for the $n=2$ model by factors 1.25 and 2 on panels (a) and (b) respectively for better visualization.
  }
  \label{fig:scaling_fun_lyapunov}
\end{figure*}

The two-point function $C_q(x,t)$ is expected (see SM \cite{SM}) to have an algebraic decay with a scaling form
\begin{equation}
    C_q(x, t) \underset{t\to\infty}{\simeq} (\lambda_q t)^{1/z_q} f_q\left(x(\lambda_q t)^{-1/z_q}\right).
\end{equation}
Here, $z_q$ is the dynamical exponent which describes the rate of spreading, and $\lambda_\alpha$ is a microscopic coefficient e.g. a diffusion constant. Finally, $f_q(s)$ is a universal scaling function whose shape is a signature of the underlying universality class.

This work mainly focuses on the transport of spin and energy at zero magnetization $\langle m^\alpha\rangle_\beta = 0$. We compute the spin and energy density correlation functions
\begin{equation}
    \begin{split}        
    C_m(x,t) &= \langle \ubm{m}(x,t)\cdot\ubm{m}(0,0)\rangle_\beta^c, \\
    C_h(x,t) &= \langle h(x,t)h(0,0)\rangle_\beta^c.
    \end{split}
\end{equation}
Here, $h(x) =\frac{1}{2n} (\partial_x \vec{m} \cdot \partial_x \vec{m})^n $ is the local energy density, and the thermal averages are taken with respect to the Gibbs measure $P[\vec{m}] \propto e^{-\beta H^{(n)}[\vec{m}]}$. 

The $n = 1$ model is the isotropic Landau-Lifschitz (LL) field theory, a classically integrable system \cite{sklyanin1979complete,takhtajan_1977_integration,faddeev_1987_hamiltonian}. Extensive numerical studies on its integrable discretization have established clear evidence of anomalous transport with a dynamical exponent $z_m = 3/2$ \cite{krajnik2020kardar,prosen2013macroscopic,ljubotina_2019_kardar,takeuchi2025partial} and the Pr\"ahofer-Spohn scaling function \cite{prahofer2004exact}. The transport of energy density in integrable systems is generically ballistic. This is due to a non-zero overlap between the energy current and an infinite number of ballistically transported conserved charges, a direct implication of the Mazur bound~\cite{mazur,zotos}. In generic non-integrable models, by contrast, conventional hydrodynamics predicts diffusive transport ($z = 2$) for both energy and magnetization \cite{spohn2014nonlinear}.

% \section{\label{sec:numerical_results}Numerical results\protect\\}
{\it Numerical results.---}
We first specify our discretization scheme to simulate the field theory numerically. The spin chain of length $L$ is placed on a lattice of $N$ sites ($L=N$ unless otherwise specified) and the spatial derivatives and their integrals are evaluated via the Fast Fourier Transform (FFT) to/from momentum space. This corresponds to imposing a momentum-space cutoff that removes the high momentum modes. In a Gibbs ensemble, such modes are already thermally suppressed and thus naturally reduce the discretization errors. 

Thermal expectation values are computed by sampling the Gibbs distribution $e^{-\beta H^{(n)}}$  through a Markov chain Monte Carlo simulation (see SM~\cite{SM}). We take a sample of the Gibbs state as an initial state and integrate the equation of motion (Eq.~\eqref{eq:dynamics}) using an adaptive-step Runge-Kutta solver with fixed tolerance. Because the Gibbs ensemble is invariant under the Hamiltonian dynamics, the system exhibits both spatial and temporal translation invariance. We exploit these symmetries to average correlation functions over space and over time, which significantly improves statistical accuracy. 

We begin by presenting the equal-space autocorrelation functions $C_q(x =0,t)$ (Fig.~\ref{fig:mm_hh_t}), from which we extract the dynamical exponents $z_q$. At infinite temperature ($\beta = 0$) the spin transport is asymptotically diffusive with $z_m=2$ in both models. For $n=1$, the energy transport at infinite temperature is also diffusive, however for $n = 2$, the energy spreads rapidly and we see signs of a diffusive central peak on top of ballistic tails (see End Matter). At finite temperature, anomalous transport behavior emerges. For both models, the spin autocorrelator decays as  $t^{-2/3}$ ($z_m = \frac{3}{2}$). The energy autocorrelators become ballistic at a much lower temperature (other field theories exhibit similar transport, see the End Matter). We expect a crossover to diffusion at long times $t^*(\beta)$, however we can observe it only at high temperatures $\beta 
\lesssim 1$.

We collapse the space-time profiles of the correlation functions in terms of their corresponding scaling variable. The spin correlator is rescaled by $x/t^{\frac{1}{z_m}}$, and we find that the correlator collapses into the universal KPZ scaling functions for both $n = 1$ and $n = 2$ models at finite temperature. Fig.~\ref{fig:scaling_fun_lyapunov} (a) demonstrates clear deviations of both distribution functions from the Gaussian and proximity to the tabulated Pr\"ahofer-Spohn scaling curves \cite{prahofer2004exact}. 

The ballistic energy transport emerges only at a much lower temperature. Fig.~\ref{fig:scaling_fun_lyapunov} (b) shows the data collapse for energy transport for $n = 1, 2$ models with the scaling variable $x/t$,  but onto different scaling functions: the $n=1$ (LL) model exhibits an oscillatory step function shape, while the $n= 2$ model has two visible peaks and a ballistic exponential tail. The energy transport is ballistic even in the non-integrable case due to momentum conservation. The current of the momentum is in the same parity sector as the energy density which leads to a non-zero energy Drude weight. The correlation profile may warrant a generalized hydrodynamics study~\cite{doyon_generalized_2025,castroalvaredo_emergent_2016,bertini_transport_2016,doyon_soliton_2018,bastianello_generalized_2018}.

The superdiffusive behavior discussed above is expected for the $n = 1$ LL model, which is a discretization of an integrable field theory. The thermal ensemble acts as a physical regulator for discretization errors. Lowering the temperature thus brings the lattice simulation closer to the underlying integrable point. The finite temperature introduces a crossover timescale under which integrable dynamics becomes visible. The End Matter shows how the extracted dynamical exponents $z_m$ approach the values found in the integrable lattice models as $\beta$ increases; for intermediate $\beta\sim 1$, we also observe a crossover to diffusive behavior.

What is surprising is that the same phenomenology also appears in the $n = 2$ model, which is believed to be non-integrable and not proximate to any known integrable model. Nonetheless, as shown in the End Matter, the dynamical exponent extracted from the spin correlator again drifts toward $z_m = \frac{3}{2}$ already when the thermal correlation length extends just a few lattice sites. 

To clarify the role of integrability in the two classes of models on the lattice, we examine FTLE in the time interval $[0,t]$, 
\begin{equation}
    \lambda(t) = \frac{1}{t} \mathrm{log}\left(\frac{\norm{\ubm{m_1}(t)-\ubm{m_2}(t)}}{\norm{\ubm{m_1}(0)-\ubm{m_2}(0)}} \right),
\end{equation}
where $\vec{m}_1$ and $\vec{m}_2$ are two nearby spin fields.  The distance is measured by the $L_2$ norm:
\begin{equation}
   \norm{\ubm{m}_1-\ubm{m}_2}^2 = {\int_{-\frac{L}{2}}^{\frac{L}{2}}| \ubm{m}_1-\ubm{m}_2|^2 \mathrm{d}x}. 
\end{equation}
To compare the two models on equal footing, we choose temperatures such that the two models have the same correlation lengths. For each model and each $\beta$, we sample an ensemble of $M+1$ thermal states $\{\mathbf{m}_i\}_{i=1}^{M+1}$. The reference state  $\mathbf{m}_1$ is generated by the Metropolis algorithm, while the other $M$ states are perturbations of $\vec{m}_1$. They are obtained by a position-dependent rotation $R(x)=\mathrm{exp}[-\varepsilon A(x)]$ where $A$ is a random skew-symmetric matrix and $\varepsilon $  is the  perturbation strength. 
\begin{figure}[t]
    \centerline{
\includegraphics[width=1.0\linewidth]{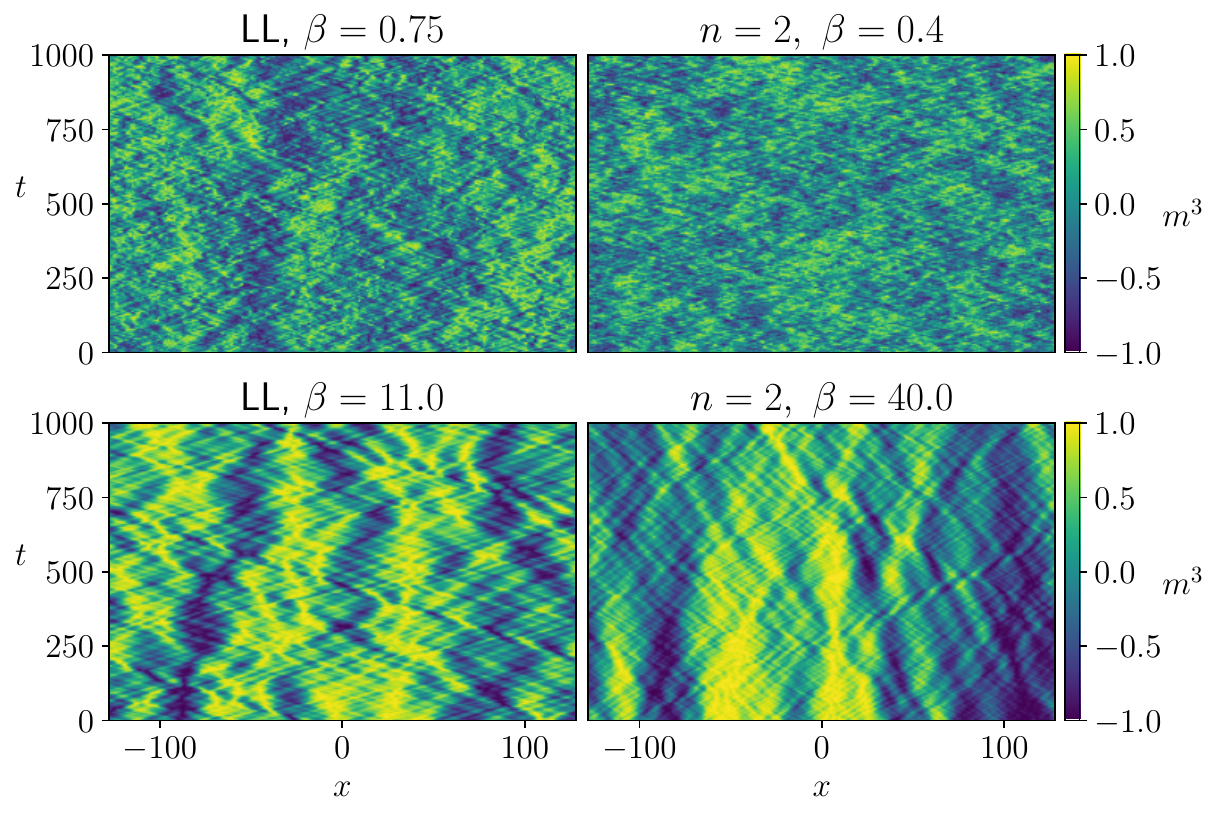}
    }
    \caption{Example of trajectories of $m^3(x,t)$ at finite temperature for the Landau-Lifschitz field theory (left panels) and the $n=2$ model (right panels). The top panels are at $\xi \approx 3a$, while the bottom panels are at $\xi \approx44a$. The gauge is fixed such that $\mathbf{m}(0,0)=(0,0,1)$. Parameters of simulation $L=N=256$, integrator tolerance $10^{-8}$.}
\label{fig:finite_beta_dynamics}
\end{figure}

Fig.\,\ref{fig:scaling_fun_lyapunov} (c) shows the results of the FTLE for both models. For thermal initial states, the dynamics of both models are clearly chaotic, as initial perturbations grow exponentially in time. However, as $\beta$ increases, FTLE of $n = 1$ model is suppressed whereas FTLE of  $n = 2$ model remains much larger. 
For the $n = 1$ model, the thermal ensemble restores approximate integrability at low temperatures, as the FTLE becomes very small. Earlier studies of integrable systems with weak integrability-breaking perturbations have shown that superdiffusive spin transport can persist up to a long time scale, which scales as a large inverse power of the integrability-breaking strength~\cite{mccarthy2024slow, mcroberts2024parametrically,wang2025breakdown}. Such superdiffusion is thus ``hard to avoid". The puzzle lies in the $n = 2$ model, as its Lyapunov exponent remains comparatively large, and the model is not known to lie near any integrable point.  

It is believed that solitons or quasi-particles in integrable models, particularly the large/`heavy' and slowly moving ones, play a central role in producing superdiffusive spin transport~\cite{gopalakrishnan_2018_kinetic,gopalakrishnan_2019_kardar,
bulchandani_2018_solvable,bulchandani_2021_superdiffusion}. We thus perform an empirical search for soliton-like excitations in our discrete models. Fig.~\ref{fig:finite_beta_dynamics} shows the spacetime heat maps of the \textit{z}-component of the spin field for both $n = 1$ and $n = 2 $ models. We choose temperatures such that the correlation lengths match in each row. In the lower panels, clear ballistic soliton-like trajectories are visible. They traverse $\sim 200$ sites over a timescale of order $10^3$. 

At higher temperatures, the spacetime profiles appear fully thermalized, with no apparent soliton-like propagation. This is consistent with the diffusive behavior observed on timescales of order $10^3$. 

The above shows the $n = 2$ model could still  be close to a hidden integrable point, which even when put on the lattice, can still host soliton-like modes within a long time scale shielded by a finite temperature. Could such an integrable point at effective low energies be just the LL model? However, this possibility is inconsistent with the observation (see SM~\cite{SM}) that fluctuations of LL conserved quantities under $n=2$ model dynamics is comparable to their thermal fluctuations at low temperatures, namely they are not conserved.

% \section{Conclusion}
{\it Conclusion.---} Integrable field theories endow continuum field theory with integrability structure. Yet the numerical investigation of their anomalous transport necessarily requires some form of discretization which typically breaks integrability. In this work, we introduce a thermal ensemble as a regulator that partially mitigates the artifacts of na\"ive discretization: even with a simple momentum space cutoff discretization, the thermal regulator enables us to observe a crossover from superdiffusive spin transport to conventional diffusion on a temperature dependent time scale. Remarkably, this crossover appears not only in the Landau–Lifshitz field theory ($n=1$) but also in the non-integrable ($n=2$) model.

Similar crossovers have been reported in earlier studies of integrable {\it lattice} models perturbed weakly away from integrability~\cite{ferreira2020ballistic, de2021stability, roy2023robustness, mccarthy2024slow, mcroberts2024parametrically, wang2025breakdown}: when the perturbation is small, the crossover time is long and superdiffusion persists over a broad time window. Our new observation is that, a generic non-integrable SO(3) symmetric field theory model (at least of the LL-type) still supports superdiffusion within a time window determined by temperature. In these models, the Lyapunov exponent is finite and no conserved quantities exist beyond total spin and total energy. Despite these chaos indicators, spin transport still contains slowly moving soliton-like trajectories~\cite{mcroberts2024parametrically}. This robustness suggests that anomalous transport should be observable experimentally in materials where exact integrability is broken, provided the relevant temperature scale is not too high compared to the emergent hydrodynamic scales.

An important direction for future work is to clarify the precise role of continuum field theory in this phenomenon. A field theory often comes with translation invariance and the associated momentum conservation. In the setup of this work, momentum conservation supports ballistic energy transport (though in a different manner in $n = 1$ and $n = 2$), but does it bring more to protect the soliton-like modes? The field theory setup without strict integrability requirement also invites a field theory calculation~\cite{de_nardis_anomalous_2019} of the transport, perhaps starting from a low temperature expansion.

\acknowledgements We would like to acknowledge Roderich Moessner, Romain Vasseur, Sarang Gopalakrishnan, Jacopo De Nardis, Vir Bulchandani, Hyunsoo Ha, Jack Kemp, and Sajant Anand for insightful discussions and/or sharing their latest results with us. We acknowledge the hospitality of the KITP program \textit{Learning the Fine Structure of Quantum Dynamics in Programmable Quantum Matter}, whose stimulating environment fostered many helpful discussions and feedback on our manuscript. This research was supported in part by grant NSF PHY-2309135 to the Kavli Institute for Theoretical Physics (KITP). TZ acknowledges Advanced Research Computing at Virginia Tech for providing computational resources.
MK and TP acknowledge support by ERC Advanced grant
  No.~101096208 -- QUEST, 
  and Research Program P1-0402 and Grant N1-0368 of Slovenian Research and Innovation Agency (ARIS).

\bibliography{refs}
\clearpage
\section{End Matter}

\textit{Appendix A: Crossover of dynamical exponents with temperature.} A general expectation for transport of magnetization for a generic model is that it will be diffusive $z_m=2$ (this can be seen in the SM \cite{SM}). However, at $\beta\sim1$
and further lowering the temperature, we observe a crossover of the dynamical exponent $z_m \to 3/2$, while the scaling function resembles that of Pr\"ahofer-Spohn.

\begin{figure}[h!]
    \centerline{
\includegraphics[width=1.0\linewidth]{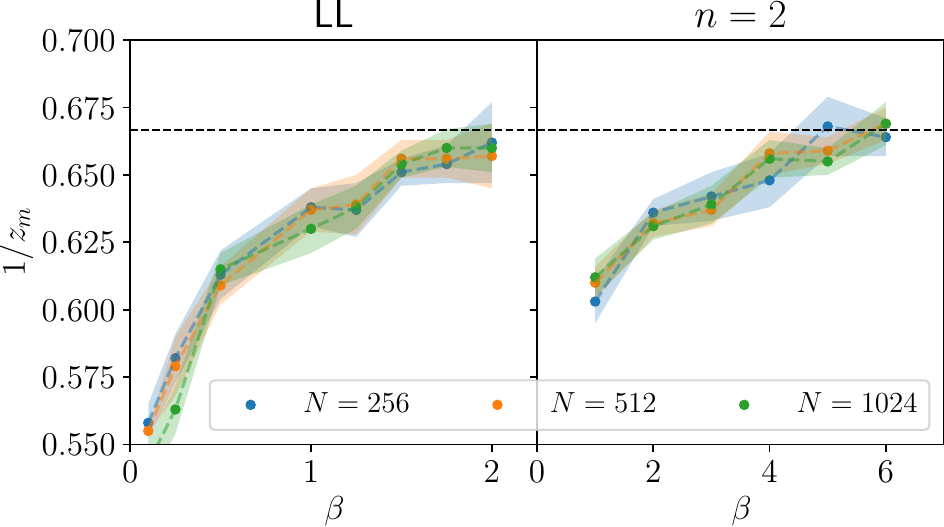}
    }
    \caption{Convergence of dynamical exponents $z_m$ for increased $\beta$. The left panel show transport in the LL field theory, while the right panel shows transport in the $n=2$ model. Shaded region shows error of fit.}
\label{fig:crossover_mm}
\end{figure}

To extract the dynamical exponent $z_m$, we fit over a finite time window up until finite-size effects are noticeable (for details see \cite{SM}). The convergence of the average dynamical exponents $z_m$ with $\beta$ is presented on Fig.\,\ref{fig:crossover_mm}. The dynamical exponent of magnetization converges to $z_m=3/2$ already at intermediate $\beta\sim1$ where the correlation length is on the order of a few lattice sites.

Strictly speaking, the dynamical exponent is a feature of a spreading peak (see SM \cite{SM}), meaning we cannot generally extract it solely from $C_q(0,t)$. This issue does not arise for magnetization transport, as the peak is stationary. The LL energy correlator has a special form with a rectangular limit-shape height $1/t$ and width $t$, as can be seen in Fig.~\ref{fig:scaling_fun_lyapunov}(b). This is a direct consequence of integrability -- there exist an infinite number of ballistic modes that overlap with the energy. We extract the dynamical exponent for this limit shape at $x=0$ which should approach $z_h=1$ if the transport of each mode is ballistic (using the same procedure as for $z_m$).

We argue that, even in the $n=2$ case, $z_h$ can be extracted from the $x=0$ contribution since the correlator collapses under a ballistic rescaling, as seen in Fig.\,\ref{fig:scaling_fun_lyapunov}(b). For $n=2$ we observe a slow convergence of $z_h(t)$ towards unity at timescales of order $t\sim 10^3$. To estimate a dynamical exponent in the $n=2$ model, we compute $z_h = \max_t z_h(t)$ for $t>10$ (a cutoff has to be used as early times do not exhibit generic behavior).

\begin{figure}[h!]
\centering
\includegraphics[width=1.0\linewidth]{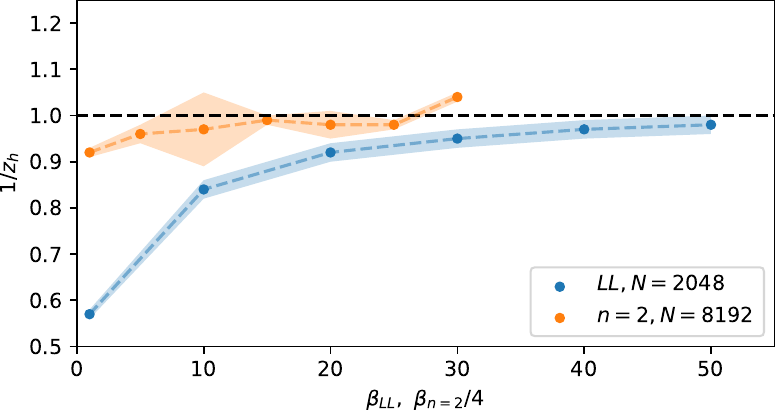}
    \caption{Convergence of the dynamical exponent for energy transport $z_h$ for increased $\beta$ in both the LL and $n=2$ models. Note the different $x$ scales for each model.}
\label{fig:crossover_hh}
\end{figure}

\begin{figure}[h!]
\centering
\includegraphics[width=0.8\linewidth]{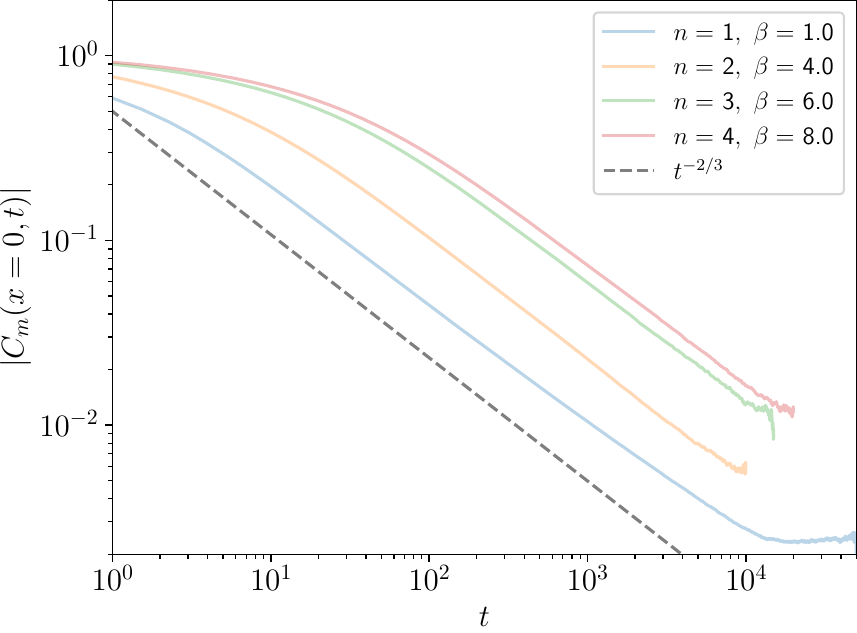}
    \caption{Dynamics of the magnetization correlator for higher $n$. The dashed black line denotes a decay with $t^{-2/3}$. Parameters of simulation $N=1024$.}
\label{fig:spin_transport_higher_n}
\end{figure}

\textit{Appendix B: Results for models with higher $n$.} 
We showcase magnetization transport for higher $n$ models. It seems that the same phenomenology applies, where $z_m=3/2$ on the observed timescales at $\beta \sim 2n$. This is presented up to $n=4$ on Fig.~\ref{fig:spin_transport_higher_n}. The $n=3$ and $n=4$ data exhibit the same apparent power-law decay as in the $n=2$ case discussed in the main text within the accessible time window.

\textit{Appendix C: Low $\beta$ energy transport.} 
At high temperature, we expect both the energy and magnetization to be diffusive eventually. This is the case for $n = 1$, and magnetization of $n = 2$ model. However, the diffusive transport of energy in the $n=2$ is not perfectly reproduced with the available system sizes. 

\begin{figure}[h!]
\centering
\includegraphics[width=1.0\linewidth]{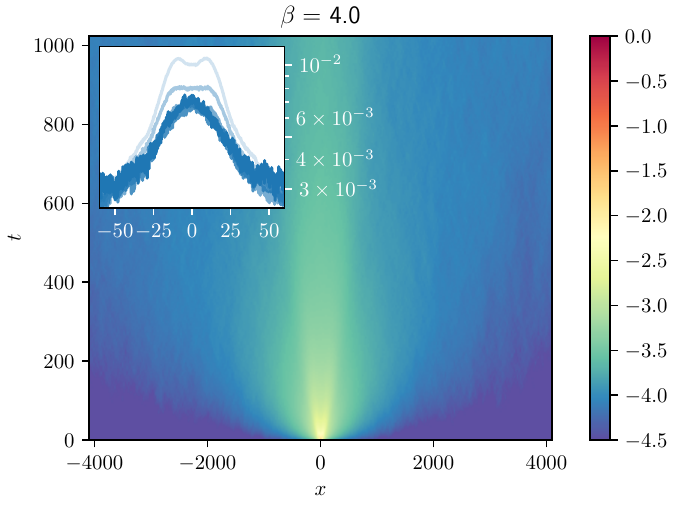}
    \caption{Heatmap of the energy correlator of the $n=2$ model at $\beta=4.0$, $N=16384$. The colors are in $\mathrm{log}_{10}$ scale. The inset shows cross-sections at increasing times from $t=125$ (light blue) to $t=1250$ (dark blue) with a diffusive rescaling $t^{1/2} C_h(x/t^{1/2},t)$ zoomed in on the central peak which becomes Gaussian.}
\label{fig:energy_cross_section_low_beta}
\end{figure}

Here we append a heatmap of the energy correlator at $\beta=4.0$ (Fig.~\ref{fig:energy_cross_section_low_beta}) to show the empirical reason. Compared with the magnetization correlator with $\beta = 0.0$, which shows clear sign of a diffusive light cone, the energy correlator spreads quickly so that the perturbation spreads to the boundary already at around $t = 25$. Therefore within our system size, we only observe the energy equilibrates due to the finite system size, rather than the free spreading into open space.

Nonetheless, we observe that at $t \gtrsim 1000 $ the central peaks collapse under a diffusive rescaling $t^{1/2} C_h(x/t^{1/2},t)$, while the exponential tails do not collapse  (see the inset on Fig.~\ref{fig:energy_cross_section_low_beta}). A larger $\beta$ helps to slow down the Lieb-Robinson light cone, and at the same time, the ballistic transport presents itself.

\textit{Appendix D: Quartic model.} In order to show more generic behavior in a non-integrable model, we show additional results of transport in the quartic theory $H=\frac{1}{2}\int \mathrm{d}x \, (\partial_x^2 \mathbf{m} \cdot \partial_x^2 \mathbf{m})$. It has the equation of motion

\begin{equation}
    \frac{\partial \mathbf{m}}{\partial t} = \mathbf{m}\times \partial_x^4 \mathbf{m}.
\end{equation}

Conveniently, both the Metropolis procedure 
\begin{figure}[h!]
 \centering
\includegraphics[width=1.0\linewidth]{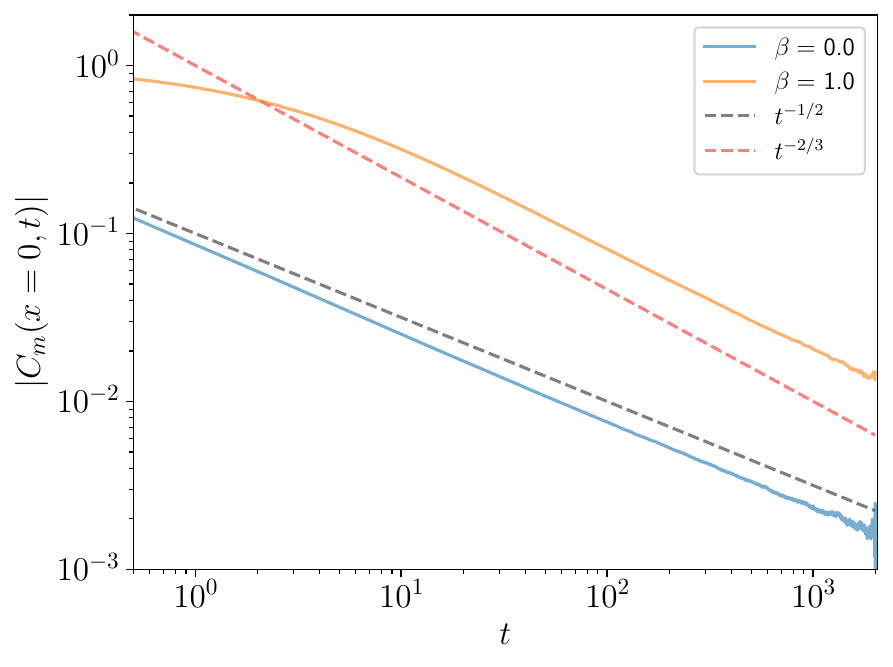}
    \caption{Dynamics of the magnetization correlator for the quartic theory at $\beta = 0.0, \ N=1024$ (diffusive) and $\beta=1, \ N=512$ (transient superdiffusive).}
\label{fig:spin_transport_quartic}
\end{figure}
and time integration can both be done in the same way as for the LL theory with only the order of the spatial derivative being larger.

The magnetization transport behaves in the same way as for the models described in the main text. Namely, at $\beta=0$, diffusion is observed, while at finite $\beta$ superdiffusive decay with $z_m=3/2$ emerges. This is shown on Fig. \ref{fig:spin_transport_quartic}.

\begin{figure}
    \centerline{
\includegraphics[width=1.0\linewidth]{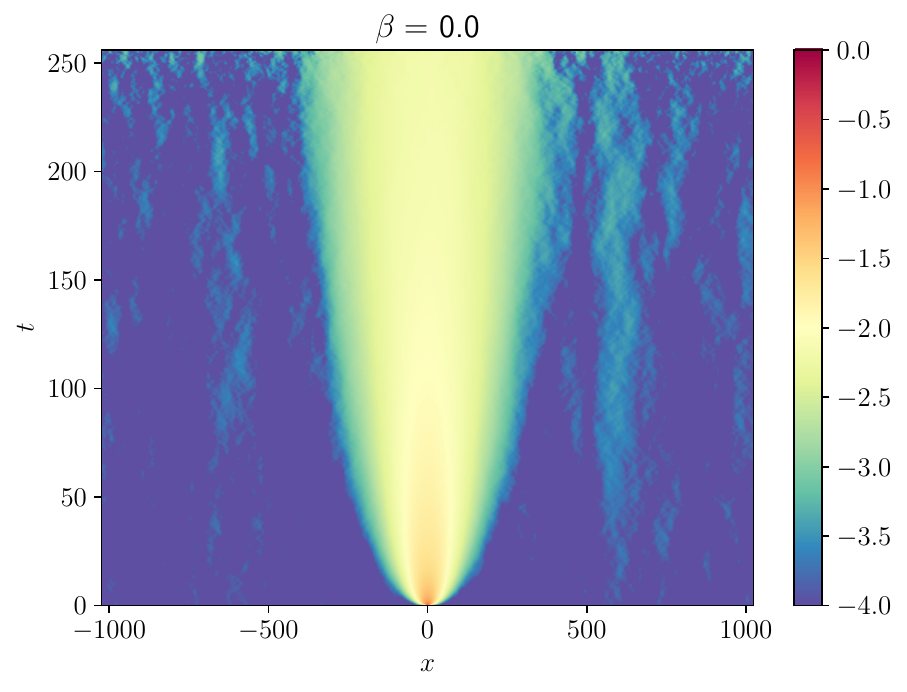}
    }
    \centerline{
\includegraphics[width=1.0\linewidth]{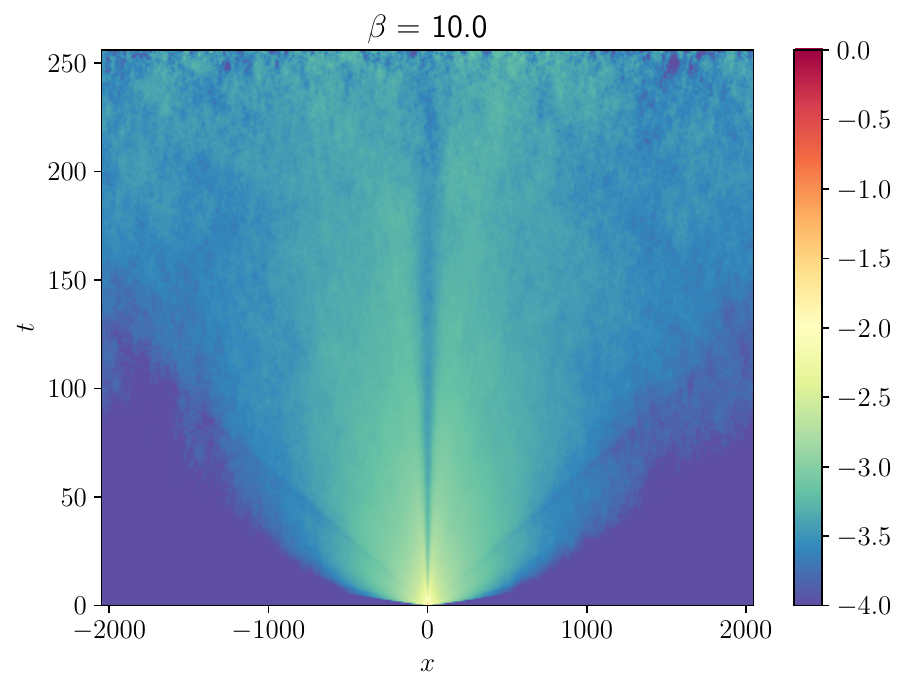}
    }
    \caption{Dynamics of the energy correlator for the quartic theory at $\beta=0$ (top) and $\beta=10$ (bottom) with $N=4096$. The high temperature regime shows purely diffusive transport, while at lower temperatures ballistic peaks become apparent. The colors are in $\mathrm{log}_{10}$ scale.}
\label{fig:energy_transport_quartic}
\end{figure}
On the other hand, energy transport exhibits more generic behavior than the $n=2$ model. The heatmaps for these two cases are shown on Fig.~\ref{fig:energy_transport_quartic}. At $\beta=0$ diffusive spreading is observed with a Gaussian profile. At $\beta=10$ there is a clearly visible ballistic light-cone which develops diffusive tails. As the tails spread the dynamics asymptotically becomes diffusive, however, the finite $\beta$ ensemble allows us to to observe ballistic transport on finite timescales at short distances from the origin.

\clearpage

\onecolumngrid
\appendix
\newpage

\begin{center}
	{\Large Supplemental Material - Anomalous transport in non-integrable classical field theories}
\end{center}
\begin{center}
	{Matija Koterle, Toma\v z Prosen, Tianci Zhou}
\end{center}

In the supplemental material, additional details regarding various technicalities are presented.

\begin{enumerate}
    \item Details of numerical calculations: we present the discretization scheme in space, along with the choice of integrator in time. The scheme is tested on the charges of the Landau-Lifschitz model using the dynamics of LL and $n=2$.
    \item Transport: discussion of the transport, along with methods used to extract the dynamical exponents.
    \item Metropolis algorithm: implementation of the Metropolis algorithm for LL, $n=2$ and quartic models.
    \item Finite-time Lyapunov exponents: details of the FTLE calculation, along with results for the continuum limit $a \to 0$.
    \item Computation of momentum: various methods for numerical evaluation of the momentum are discussed, which is conserved in the continuum limit.
\end{enumerate}

\section{Details of numerical calculations}

A na\"{i}ve spatial discretization of an integrable field theory will break integrability \cite{krajnik2020kardar}, since the tower of charges will generally not be conserved during evolution of the differential-difference equation of motion produced by the discretization procedure. To preserve integrability one must consider integrable discretizations \cite{ishimori1982integrable, haldane1982excitation}. For the purposes of numerical simulations, integrable discretizations can be made even in time (see \cite{date1982method} for a general overview or \cite{krajnik2020kardar} for the LL model) -- such difference-difference dynamics can be thought as special integration schemes. For non-integrable equations of motion it is not clear how to construct discretized versions that would preserve the same physical properties. We now discuss the discretization scheme employed in this work.

\noindent
\textbf{Discretization.} The periodic domain $x\in(-L/2, L/2]$ is split into $N$ equidistant points with lattice spacing $a=L/N$. We approximate integrals with Riemann sums $\int_{-L/2}^{L/2} \mathrm{d}x f(x)= a \sum_{i=1}^N f(x_i) + \mathcal{O}(a^2)$, while derivatives in $x$ are computed in Fourier space. This is done by calculating $\mathbf{m}(x) \approx \sum_{k=-N/2}^{N/2-1} e^{-\mathrm{i} \frac{2\pi x}{L}} \mathbf{m}_k$ and applying the derivative operator $(-\frac{2\pi}{L} \mathrm{i}k)$.

The discretization also affects the Gibbs measure, namely, one must take the correct field-theory scaling limit. When $N$ is doubled, the inverse temperature $\beta$ is halved, flowing to $\beta = 0$ in the continuum limit $a \to 0$. Our protocol is to double $N$ and $L$ at a fixed $a$ which keeps the energy scale $\beta$ equal while diminishing finite-size effects, flowing to the thermodynamic limit $L\to\infty$ of the field theory.

\noindent
\textbf{Numerical integration.} To compute dynamics we must numerically evaluate the equation of motion
\begin{equation}
    \frac{\partial\mathbf{m}}{\partial t} = - ~ \mathbf{m}\times \partial_x[(\partial_x \mathbf{m} \cdot \partial_x\mathbf{m})^{n-1} \partial_x\mathbf{m}].
\end{equation}
To do this we use a pseudo-spectral approach to compute derivatives in Fourier space as described above. This splits the PDE into a system of coupled ODEs for the $k-$modes. The final step is to discretize time. This is done in \verb|Julia| with the package \verb|DifferentialEquations.jl| \cite{rackauckas2017differentialequations}, making use of Verner's "most efficient" 6th order Runge-Kutta method \cite{verner1978explicit}. This is an adaptive Runge-Kutta algorithm where the time-stepping is set by an explicit choice of trajectory tolerance. Unless stated otherwise we always used a tolerance of $10^{-8}$. This (discretized) form should manifestly preserve the norm $|\mathbf{m}(x)|=1$, and the total magnetization $\mathbf{M}=\int \mathrm{d}x\, \mathbf{m}(x)$.

\begin{figure}[h!]
    \centerline{
    \includegraphics[width=1.0\linewidth]{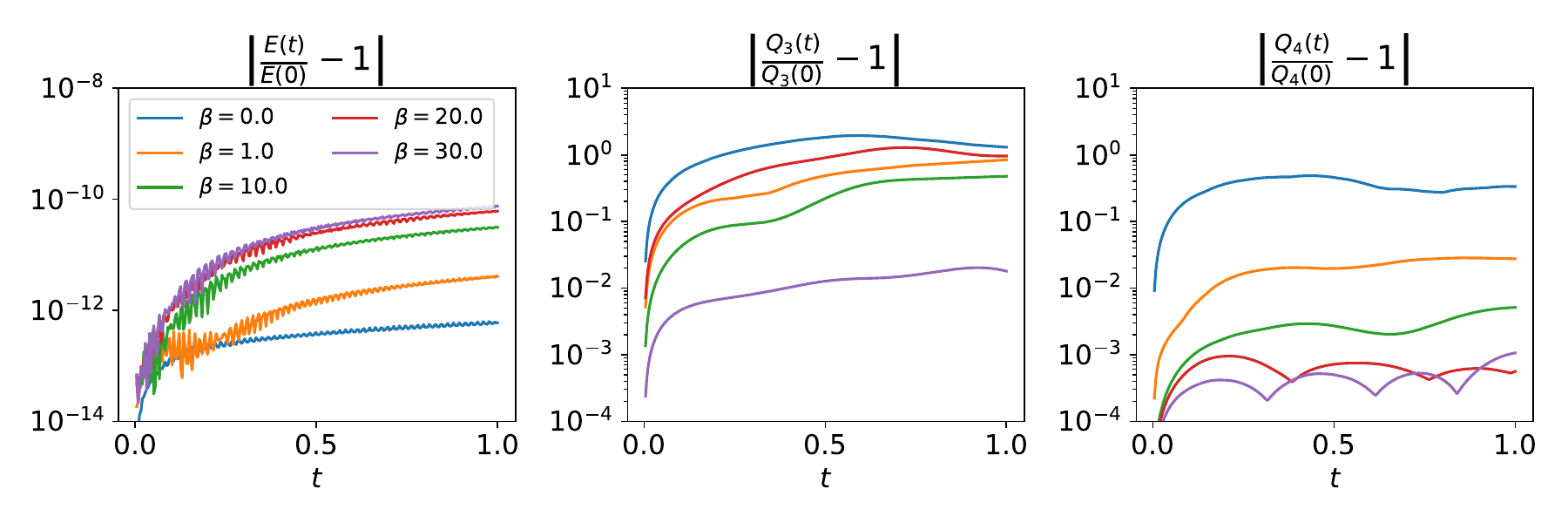}
    }
    \caption{Relative conservation of charges in LL at increased $\beta$ of a single trajectory. The panels show $E$, $Q_3$ and $Q_4$ from left to right (temperatures are the same as on the left plot). The lines show the mean over over 10 trajectories. Parameters of simulation $N=L=1024$, integrator tolerance $10^{-10}$.}
\label{fig:conservation}
\end{figure}

\noindent
\textbf{Conservation of charges.} We display the conservation of some charges during numerical integration of thermal states at increased $\beta$. The charge we consider for both models is the energy. For the LL field theory, the next two charges in the tower of conserved quantities are
\begin{equation}
    \begin{split}   
        &Q_3 = \int \mathrm{d}x \, \mathbf{m}\cdot(\partial_x\mathbf{m}\times\partial_x^2\mathbf{m}) = - \int \mathrm{d}x \, \partial_x\mathbf{m}\cdot(\mathbf{m} \times\partial_x\mathbf{m})_x,\\
        & Q_4 = \frac{1}{2} \int_{-L/2}^{L/2}\mathrm{d}x\, \left[ (\partial_x^2\mathbf{m}\cdot\partial_x^2\mathbf{m})-\frac{5}{4}( \partial_x\mathbf{m}\cdot \partial_x\mathbf{m})^2 \right].\\
    \end{split}
\end{equation}
\noindent
The $n=2$ model does have additional conserved quantities, namely the magnetization and momentum, but the former is conserved by the integrator while the latter will be discussed in a later section. The results on Fig.\,\ref{fig:conservation} show that an increase in $\beta$ leads to smaller relative changes of the charge during time evolution.

\noindent
\textbf{Conservation of Landau-Lifschitz charges in $n=2$ dynamics.} In order to investigate vicinity to an integrable point, we compute dynamics of the LL charges using the equation of motion of the $n=2$ model. This effectively measures how close the $n=2$ flow is to the LL tower of charges. The results show that the charges fluctuate on the same scale as fluctuations in the ensemble $e^{-\beta H^{(2)}}$, suggesting the conservation does not significantly improve up to $\beta=10$.

\begin{figure}[h!]
    \centerline{
    \includegraphics[width=1.0\linewidth]{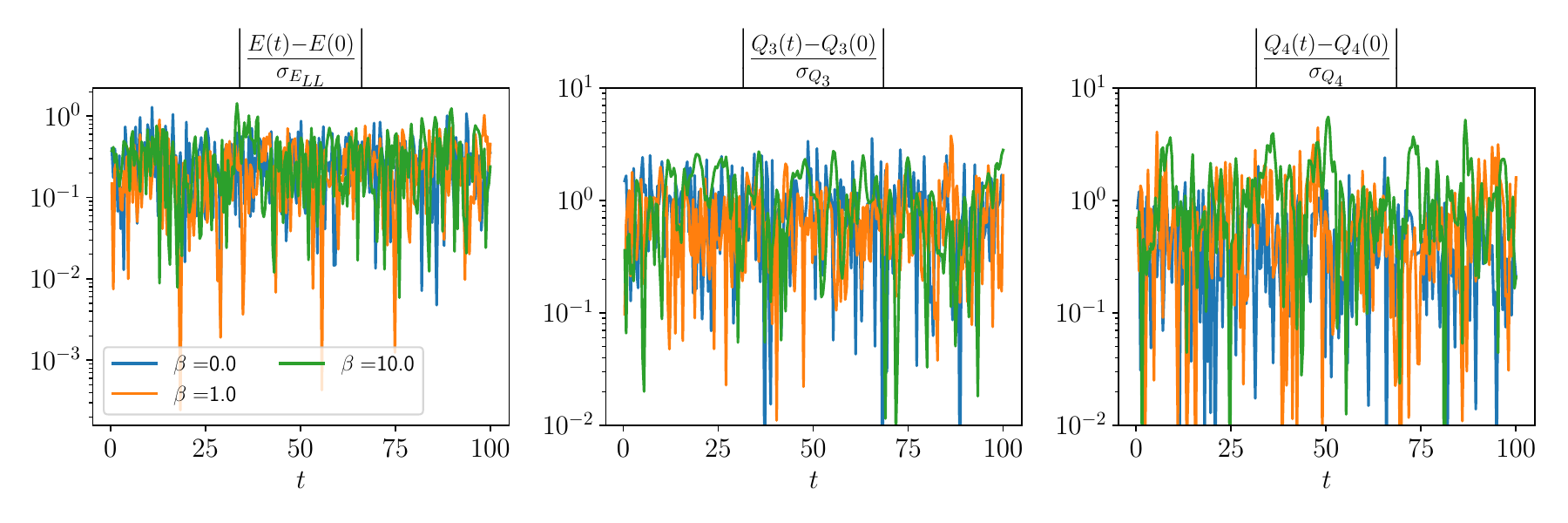}
    }
    \caption{Relative conservation of LL charges in with $n=2$ dynamics with increased $\beta$, compared to ensemble fluctuations $\sigma_q$ of the $n=2$ model. The panels show $E$ (Landau-Lifschitz energy!), $Q_3$ and $Q_4$ from left to right. Parameters of simulation $N=1024$, integrator tolerance $10^{-10}$.}
\label{fig:conservation_n_2}
\end{figure}

\newpage

\section{Transport}
In this section we will discuss some details pertaining to transport, and how to characterize different types of transport numerically. By the scaling hypothesis, asymptotically (as defined below), two-point functions of densities of conserved quantities are expected to be of the following form
\begin{equation}
    C_q(x, t) \simeq \sum_{\alpha} \frac{\chi_\alpha}{(\lambda_\alpha t)^{-1/z_\alpha}} f_\alpha\left( \frac{x-c_\alpha t}  {(\lambda_\alpha t)^{1/z_\alpha} }\right),
\end{equation}
as $(x-c_\alpha t) (\lambda_\alpha t)^{-1/z_\alpha} \to \infty$. We decompose the correlator into contributions from several normal modes labeled by $\alpha$ with weights $\chi_\alpha$. Each normal mode is characterized by $\{c_\alpha, z_\alpha, \lambda_\alpha\}$. Here, $c_{\alpha}$ is the mode velocity which denotes if a mode is ballistic ($c_\alpha\neq0$) or not ($c_\alpha=0$). Furthermore, $z_\alpha$ is the dynamical exponent which describes the rate of spreading of each mode. We stress that a ballistic mode can have an intrinsic $z_\alpha$ which is not necessarily equal to unity. The coefficient $\lambda_\alpha$ is a diffusion constant (transport coefficient). Finally, $f_\alpha(s)$ is a universal scaling function whose shape is a signature of the underlying universality class. Note that the mode decomposition can be subject to symmetry constraints (in our case, for example, it must satisfy parity symmetry). Such decompositions are standard in non-integrable models as the number of normal modes is exactly the number of conserved quantities. Consequently, in integrable models, the number of conserved quantities is extensive and this sum becomes infinite.

Although we extract the dynamical exponent from the decay of the correlator at $x=0$ (as described below), this fails if one wishes to extract $z_\alpha$ for a ballistic peak. Strictly speaking, the most robust way to extract $z_\alpha$ is from the variance growth
\begin{equation}
    \sigma^2_q(t) \equiv \int \mathrm{d}x x^2 C_q(x,t) \underset{t \to \infty}{\simeq} \mathcal{D} t^2 + L t^{2/z_\alpha},
\end{equation}
\noindent
where the leading contribution is quadratic due to the ballistic peak ($\mathcal{D}$ is the Drude weight) while the spreading acts as the first correction ($L$ is the transport coefficient). If a mode is not ballistic ($\mathcal{D}=0$) the leading contribution is on the order $t^{2/z_\alpha}$, allowing for the extraction of $z_\alpha$. In practice this definition is not very useful, especially when dealing with noisy data, so we do not use it in numerical computations.

\noindent
\textbf{Correlation functions} are calculated by using a two-dimensional Fourier transform. The transform is cyclic in the spatial direction, as we consider periodic boundaries. On the other hand, in the time direction one has to remove circular effects by padding the data and correctly renormalizing the correlator with the size of the window.

\begin{figure}[h!]
    \centerline{
    \includegraphics[width=0.5\linewidth]{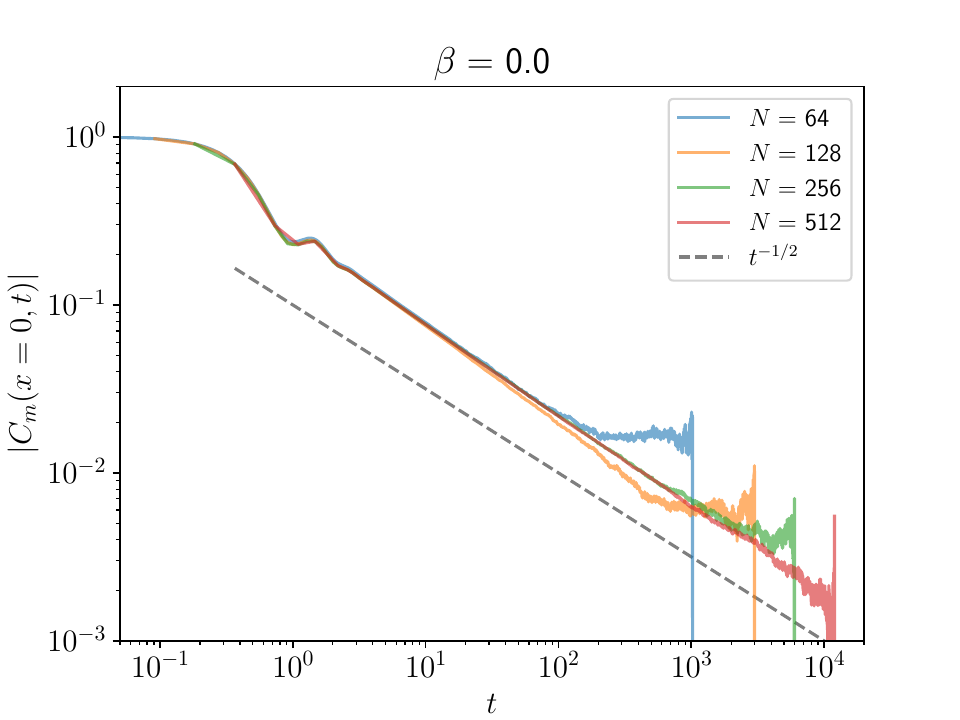}
    \includegraphics[width=0.5\linewidth]{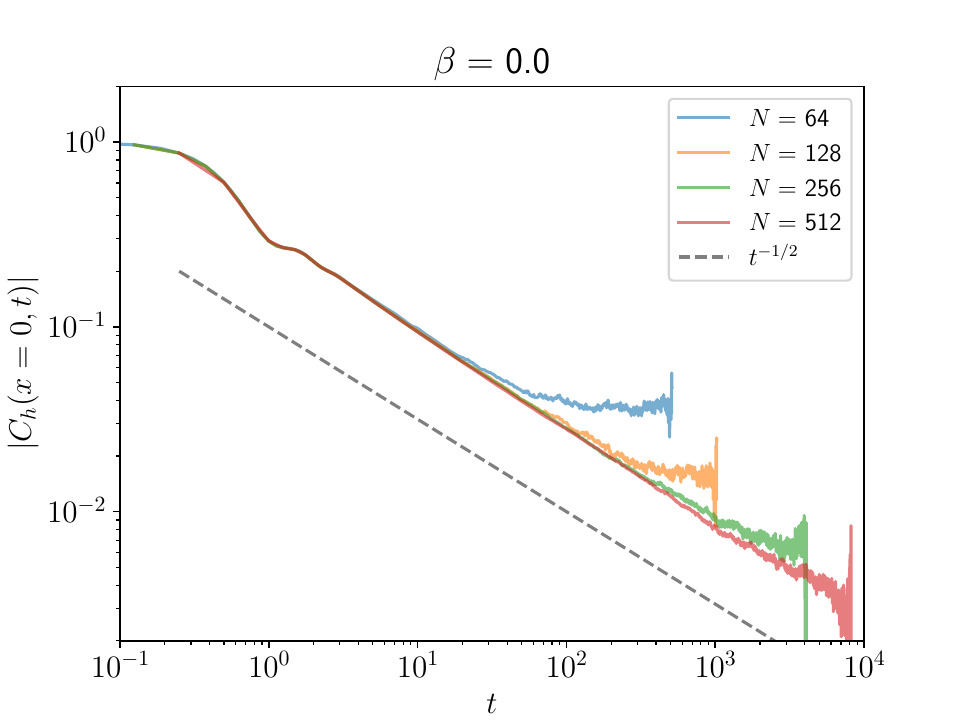}
    }
    \caption{Transport at $\beta=0$ and $N=L$ in the LL ($n=1$) model. \textbf{Left.} Decay of the $SO(3)$ invariant magnetization autocorrelation function. \textbf{Right.} Decay of the energy density autocorrelation function. The dashed lines show diffusive decay $t^{-\frac{1}{2}}$.}
\label{fig:inf_temp_LL}
\end{figure}

\noindent
\textbf{Infinite temperature.} We show an example of the infinite temperature case $\beta=0$, where all $\mathbf{m}(x_i)$ are drawn randomly from the unit-sphere. The integrator preserves only magnetization and the norm manifestly, other conserved quantities are free to drift. The observed dynamical exponent $z$ is close to $z=2$, indicating diffusive transport asymptotically as shown in Fig.~\ref{fig:inf_temp_LL}.  The finite size effects manifest themselves in plateaus of the autocorrelation functions, which are suppressed when $L$ is increased. They are related to the variances
\begin{equation}
    C_m(x, \infty) = \frac{\langle \mathbf{M}^2\rangle}{L^2}, \quad C_h(x, \infty) = \frac{\langle E^2\rangle - \langle E\rangle^2}{L^2},
\end{equation}
\noindent
where $\mathbf{M}=\int\mathrm{d}x \, \mathbf{m}(x)$.

\noindent
\textbf{Numerical evaluation of dynamical exponents.} The dynamical exponent can be computed as a log derivative of the two-point function
\begin{equation}
    z(t) = \frac{\mathrm{d} \mathrm{log} (C(0,t))}{\mathrm{d}\mathrm{log}(t)}, \qquad z=\lim_{t\to\infty} z(t).
\end{equation}
\noindent
In practice this definition is too noisy to be useful. We make use of two approaches

\begin{enumerate}
    \item Approximation of the log derivative by fitting windows of $C(0,t)$ in time. This has the effect of averaging out noise within a window, giving a more accurate local $z(t)$ than the log derivative.
    \item Fitting the entire $C(0,t)$ with a power law fit. This is systematically wrong, but as it approaches the correct results when discretization effects are small. This is done after $z(t)$ converges to a maximal value, but before the finite size plateau is reached. If fits are performed in a different window of time, the window will be stated in the text.
\end{enumerate}

\begin{figure}[h!]
    \centerline{
    \includegraphics[width=1.0\linewidth]{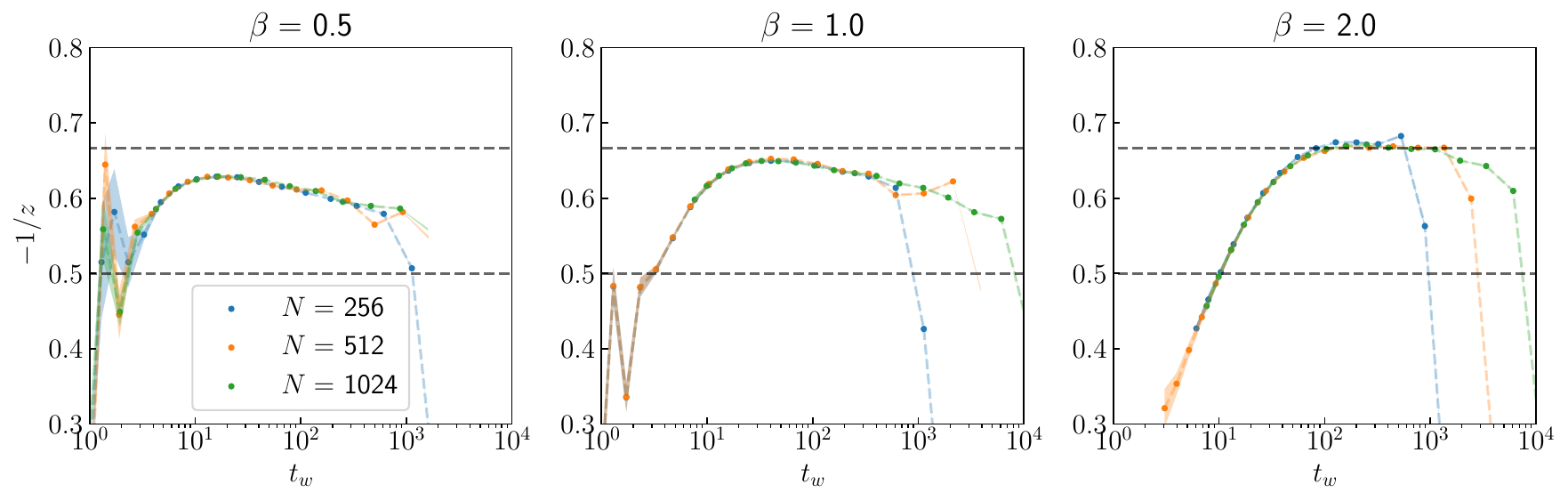}
    }
    \caption{Dynamics of $z(t)$ for magnetization transport at increased $\beta$ for the Landau-Lifschitz model. Fitting was performed in windows centered at $t_w$ (their widths can be read off from this plot).} 
\label{fig:3_plot}
\end{figure}

\noindent
Method (1.) is used to compute $z(t)$, an example is shown for the magnetization transport in the LL model on Fig.~\ref{fig:3_plot}. The estimation of $z(t)$ is robust to changing the window size used for local fits so long as the number of samples is sufficient. The drift towards $z=2$ that is observed at $\beta=0$ is suppressed with increased $\beta$. For the LL example the energy correlator has rapid oscillations as shown below. This makes method (1.) largely ineffective.

\begin{figure}[h!]
    \centerline{
\includegraphics[width=1.0\linewidth]{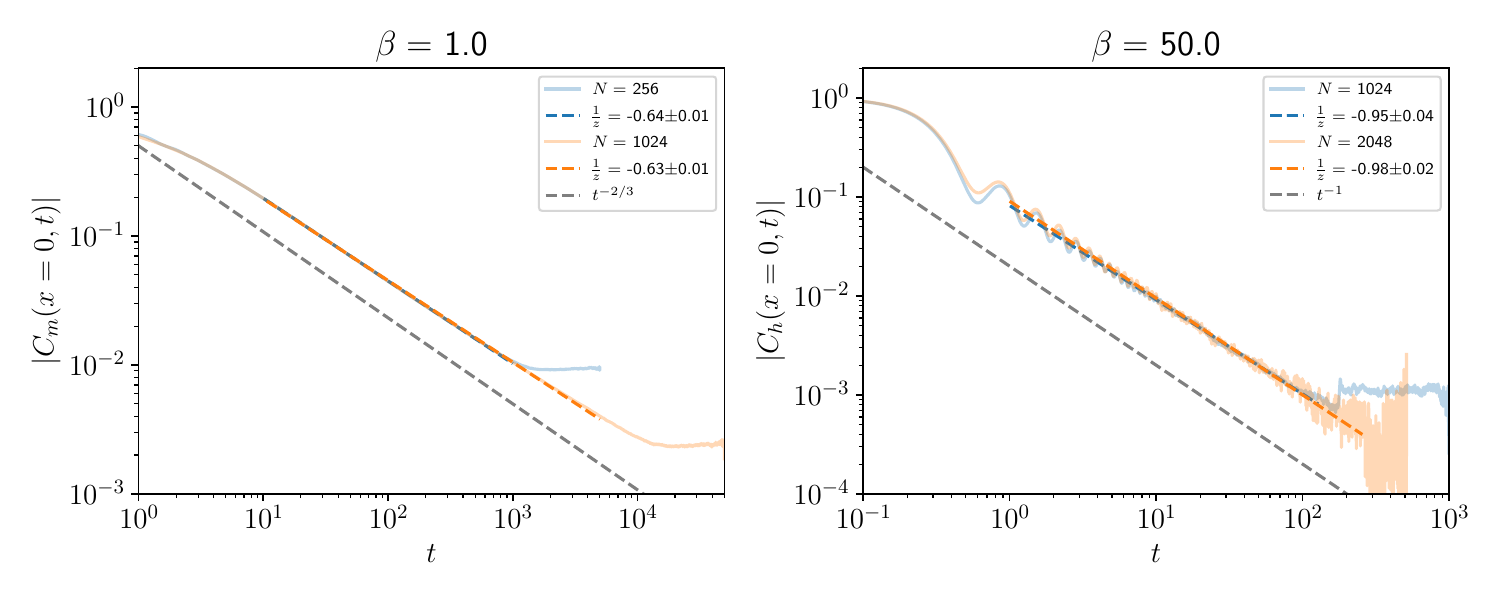}
    }
    \caption{Example of fitting the average dynamical exponent (method 2) for the LL model. \textbf{Left.} Shows fitting for magnetization transport at  $\beta=1.0$. \textbf{Right.} Shows fitting for energy transport at $\beta=50.0$.} 
\label{fig:method_2}
\end{figure}

The method (2.) is used to compute an approximate final $z=\lim_{t\to\infty}z(t)$. In all of our simulations it is expected that, given long enough, $z=2$ due to discretization effects even for non-integrable models with superdiffusive transport. By fitting $C(0,t)$ over the entire region that was computed, barring finite size effects, we effectively obtain the average $z$. An example of this for the LL model is shown on Fig.~\ref{fig:method_2}.

\newpage
\noindent
\textbf{Examples for $n=2$.} We now showcase some examples of transport of both magnetization and energy for the $n=2$ model.

\begin{enumerate}
\renewcommand{\labelenumi}{(\roman{enumi})}
    \item \textbf{Magnetization transport} exhibits a similar phenomenology as the LL case. Increasing $\beta$ leads to an increase of the dynamical exponent towards $z=3/2$. The example on Fig.\,\ref{fig:magnetization_transport} shows examples for $\beta=0.25$ and $\beta=1.0$. During evolution magnetization is preserved exactly up to numerical accuracy.
    
\begin{figure}[h!]
    \centerline{
\includegraphics[width=0.5\linewidth]{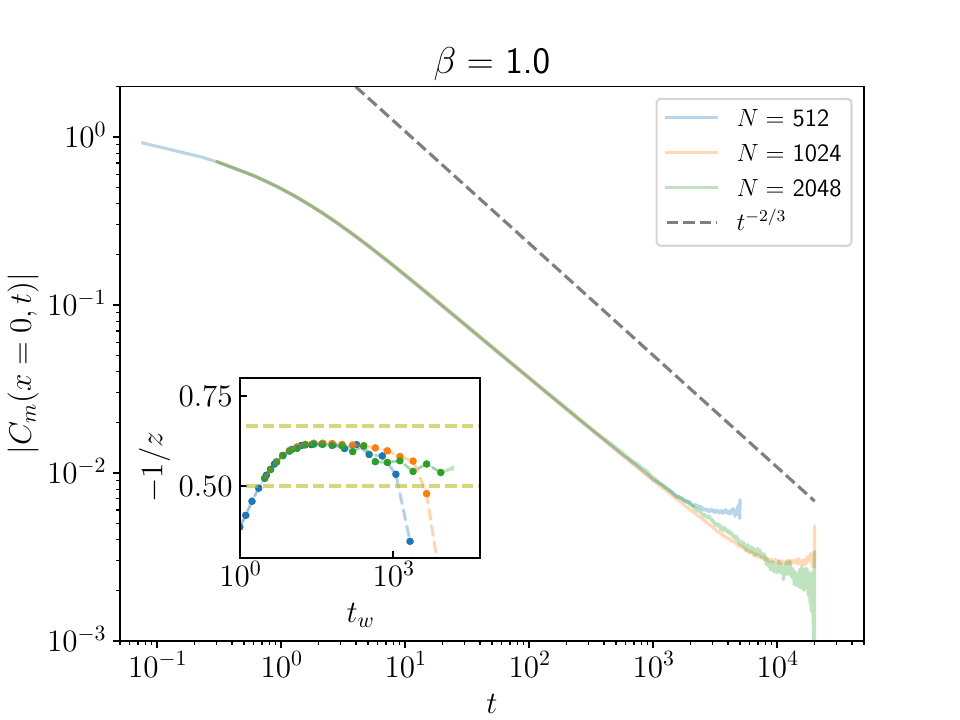}
\includegraphics[width=0.5\linewidth]{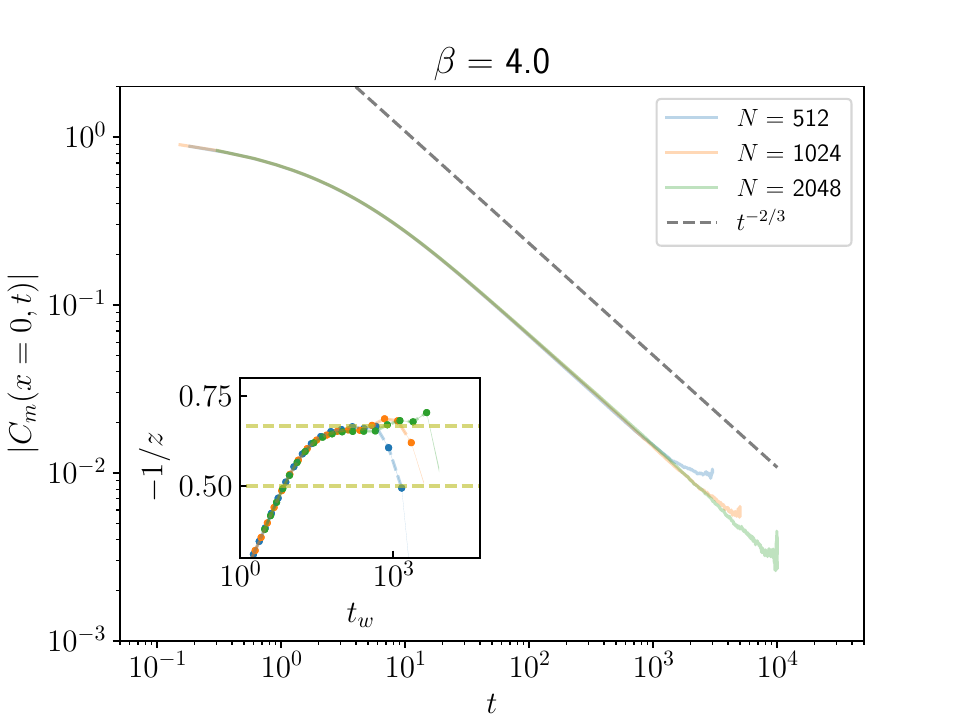}
    }
\caption{Magnetization transport in $n=2$ for $\beta=1.0$ (left) and $\beta=4.0$ (right). The insets show segmented fits of the correlator giving $z(t)$.}
\label{fig:magnetization_transport}
\end{figure}

    \item \textbf{Energy transport.} In comparison with LL the correlator loses the rapid oscillations. We measure the decay of the ballistic peak with positive velocity which shows slow dynamics of $z(t)$ that converges to $z_h \approx 1$.

\begin{figure}[h!]
    \centerline{
\includegraphics[width=0.5\linewidth]{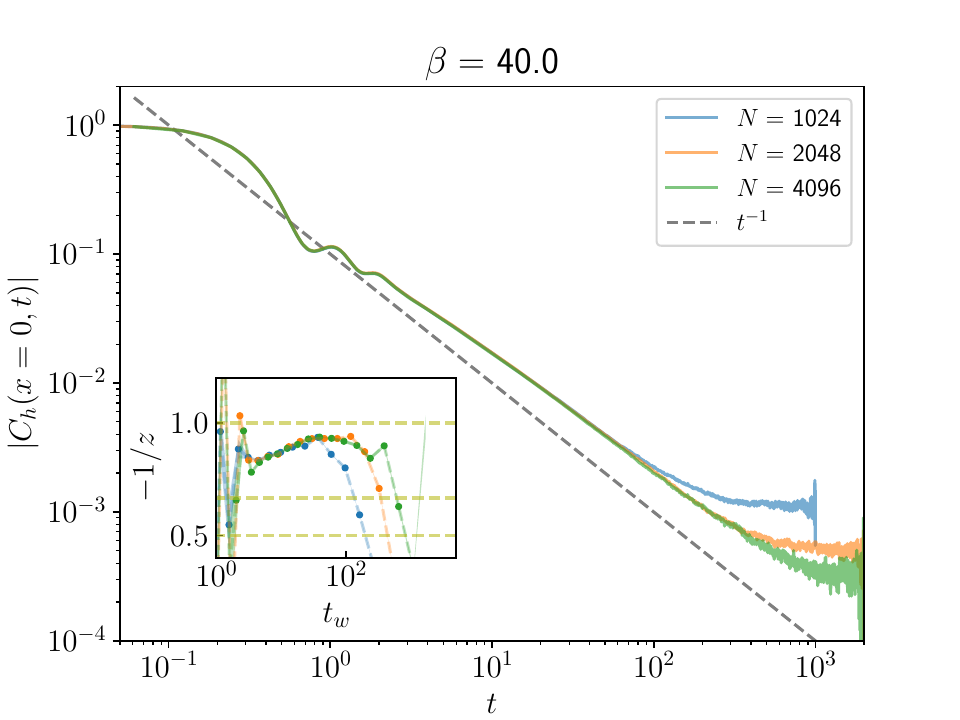}
\includegraphics[width=0.5\linewidth]{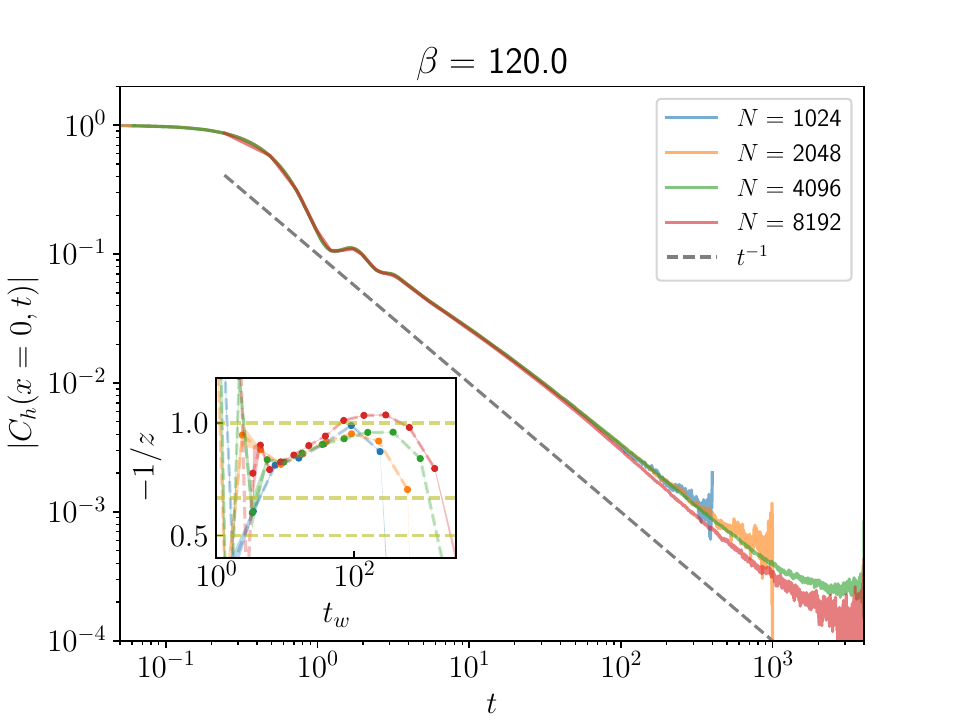}
    }
    \centerline{
\includegraphics[width=0.5\linewidth]{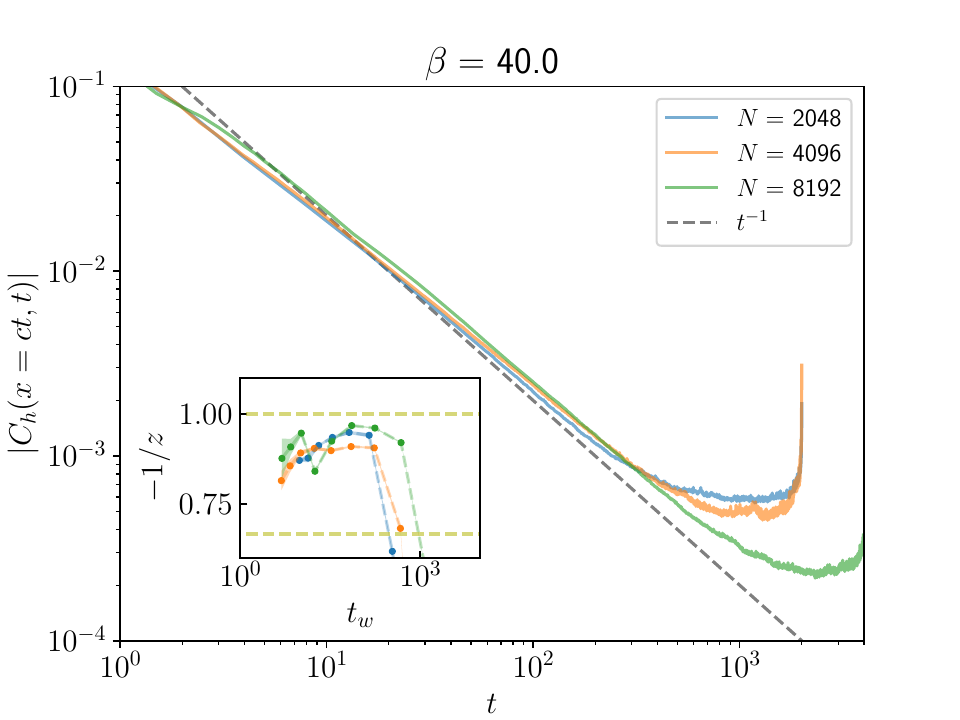}
\includegraphics[width=0.5\linewidth]{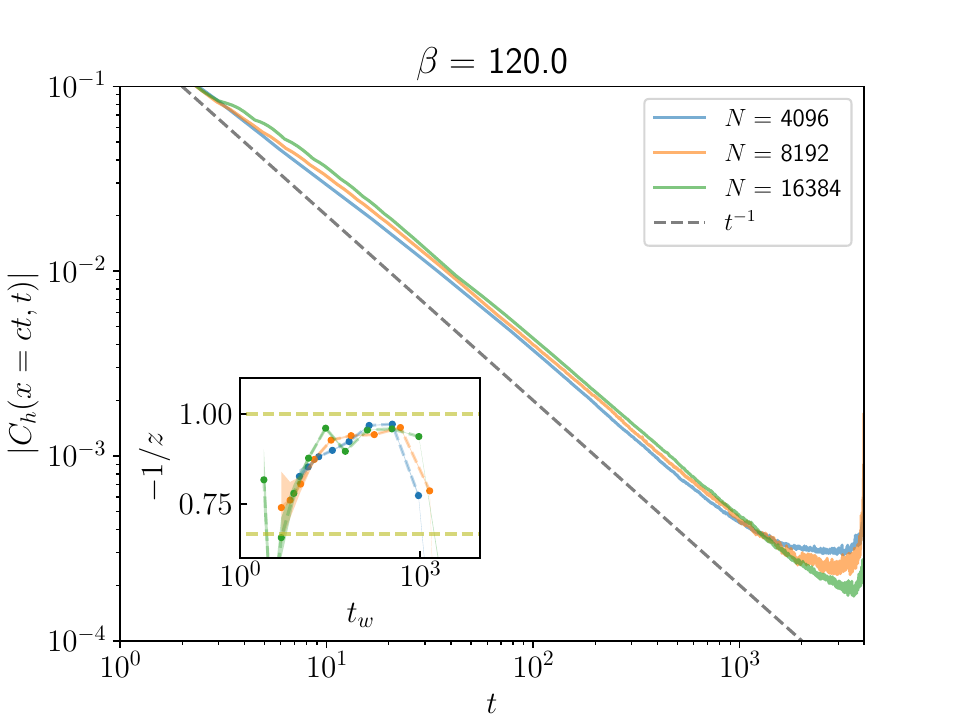}
    }
    \caption{Energy transport in $n=2$ for $\beta=40.0$ (left) and $\beta=120.0$ (right) at $x=0$ (top panels) and $x=ct$ (bottom panels). The insets show segmented fits of the correlator giving $z(t)$. Parameters of simulation: $L=N$, integrator tolerance $10^{-8}$.} 
\label{fig:energy_transport}
\end{figure}
\end{enumerate}

\clearpage
\section{Metropolis algorithm}\label{appendix:metropolis}
\noindent
To sample states from the Gibbs measure we make use of the Metropolis algorithm. In each step of the algorithm the proposed update is generated by sampling a random unit-length vector $\mathbf{m}'$ uniformly on the sphere, and an update site $x_i$ uniformly from the set $\{x_1, ... x_N\}$. The proposed update is then $\mathbf{m}(x_i) \to (1-\alpha)\mathbf{m}(x_i) + \alpha \mathbf{m}'(x_i)$, with additional norm preservaton. The update is accepted with a probability $\mathrm{min}\left(1, e^{-\beta H^{(n)}}\right)$. 

\noindent
\textbf{Local algorithm}. We first discuss a special case of the Metropolis algorithm for the Landau-Lifschitz ($n=1$) case. When a Metropolis update $\mathbf{m} \rightarrow \mathbf{m}+\Delta\mathbf{m}$ is proposed, the energy difference (of the discretized field theory with lattice spacing $a$)
\begin{equation}
    \begin{split}
    \Delta E &= -\frac{a}{2} \sum_{j=0}^{N-1}  \left[\partial_x^2\Delta\mathbf{m}(j) \cdot \mathbf{m}(j) + \partial_x^2\mathbf{m}(j) \cdot \Delta \mathbf{m}(j) + \partial_x^2 \Delta\mathbf{m}(j) \cdot \Delta \mathbf{m}(j)   \ \right] \\
    &=  -\frac{a}{2} \sum_{j=0}^{N-1}  \left[\partial_x^2\Delta\mathbf{m}(j) \cdot \Delta \mathbf{m}(j) + 2 \partial_x^2 \mathbf{m}(j) \cdot \Delta \mathbf{m}(j) \right] \\
    &= -\frac{a}{2} \sum_{j=0}^{N-1} \Delta \mathbf{m}(j) \cdot \left[\partial_x^2\Delta\mathbf{m}(j)   + 2 \partial_x^2\mathbf{m}(j) \right] \\
    \end{split}
\end{equation}
\noindent
where in the second step we integrate by parts twice to move the differential operator from $\Delta\mathbf{m}$ to $\mathbf{m}$. Note that this operation is only allowed inside the sum. The discretization of the field $\mathbf{m}$ can be represented as a matrix of size $(3,N)$. If the Metropolis update is local (only one site will be updated), say at $j=s$
\begin{equation}
    \Delta E = -a \Delta \mathbf{m}(s) \cdot \left[\frac{1}{2}\partial_x^2\Delta\mathbf{m}(s)   + \partial_x^2 \mathbf{m}(s) \right].
\end{equation}
\noindent
For efficient numerical evaluation, one can pre-compute $\mathbf{v}:=\partial_x^2 \underbrace{(1,0,0...)}_{N}$ in Fourier space and use it during the Metropolis routine. This can be used to compute $(\Delta \mathbf{m})_{xx}(s)$ by linearly shifting the vector $\mathbf{v}$ $s$ times (we denote this $\mathbbm{T}^s \mathbf{v}$)  and multiplying element-wise with $\Delta \mathbf{m}(s)$ (which is a vector of length 3)
\begin{equation}
    \left[\frac{1}{2}\partial_x^2\Delta\mathbf{m}(s)   +  \partial_x^@\mathbf{m}(s) \right] = \left[\frac{1}{2} (\mathbbm{T}^s\mathbf{v}) \cdot \Delta\mathbf{m}(s)   +  \partial_x^2 \mathbf{m}(s) \right]
\end{equation}
\noindent
The additional cost to the algorithm is that we need to keep track of $\partial_x^2\mathbf{m}$ during the procedure. Notably, the shift can be done \textbf{after} calculating the energy difference. This makes checking $\Delta E$ cost $O(1)$. In the case of an accepted we must update the field $\mathbf{m}$ and its second derivative at coordinate $s$, which costs $O(N)$
\begin{equation}
        \mathbf{m}(s) \Rightarrow \mathbf{m}(s) + \Delta\mathbf{m}(s), \quad         \partial_x^2\mathbf{m}(s) \Rightarrow  \partial_x^2\mathbf{m}(s)+(\mathbbm{T}^s\mathbf{v}) \Delta \mathbf{m}(s).
\end{equation}

\noindent
The same procedure can be applied to the quartic model with $h(x) = \frac{1}{2} \partial_x^2\mathbf{m}\cdot\partial_x^2\mathbf{m}$.

\noindent
\textbf{General algorithm.} For $n>1$ we cannot perform the same efficient update procedure due to the higher powers of the energy functional. In this case we make use of an $O(N)$ evaluation of $\Delta E$. We still pre-compute the vector $\mathbf{v}$, $\mathbf{m}_x$ and the energy $E$. The difference is that when an update is proposed at site $s$ we compute
\begin{equation}
    \partial_x \mathbf{m} \to \partial_x \mathbf{m} + (\mathbbm{T}^s \mathbf{v}) \partial_x\mathbf{m} := \partial_x \mathbf{m}'.
\end{equation}
\noindent
The new energy is then computed by definition
\begin{equation}
    E'  = a \sum_{x_i}  (\partial_x \mathbf{m}' \cdot \partial_x \mathbf{m}')^n,
\end{equation}
and $\Delta E = E' - E$ with a cost of $O(N)$ at every Metropolis step.

\noindent
\textbf{Participation ratio.} The \textit{participation ratio} (PR) is a proposed estimator for the correlation length $\xi$. We define the correlation function $C(x,t):=\langle A(x,t)A(0,0)\rangle$ which we first normalize $p(x,t):= C(x,t) / \int\mathrm{d}x \, C(x,t)$. This can be used to define the PR
\begin{equation}
    \mathrm{PR}^{-1}(t) := \int \mathrm{d}x \, p^2(x,t).
\end{equation}
\noindent
We note that all integrals are numerically evaluated as Riemann integrals. The PR can be used to benchmark the Metropolis algorithm by tracking the convergence of the correlation length at a given $\beta$. Note that this timescale differs significantly from the convergence of one-point observables, e.g. average energy.

\newpage
\section{Finite-time Lyapunov exponents}
To confirm the nonintegrability of the $n=2$ typically the Lyapunov spectrum would be computed using a variant of Benettin's algorithm \cite{benettin1980lyapunov}. However, due to the na\"ive discretization of the equation of motion, even an integrable PDE will asymptotically produce a non-zero maximal Lyapunov exponent.

Thus we instead opt to extract chaotic dynamics from short-time behavior of the system, which can show features of integrable motion. To do this we use the finite-time Lyapunov exponent (FTLE), which measures how fast the trajectories deviate in finite time, over a time interval of length $T$. It is defined as
\begin{equation}
    \lambda_{FT}(T) = \frac{1}{T} \mathrm{log}\left(\frac{\||\mathbf{m_1}(x,t)-\mathbf{m_2}(x,t)\|}{\|\mathbf{m_1}(x,0)-\mathbf{m_2}(x,0)\| } \right),
\end{equation}
\noindent
where $\mathbf{m}_1=\mathbf{m}_1(x,t)$ is the reference trajectory and $\mathbf{m}_2=\mathbf{m}_2(x,t)$ is a perturbed trajectory close to the initial one. If the two trajectories deviate exponentially, then the motion is chaotic in this region of phase space.

\textbf{Norm.} Suppose we have two configurations $\mathbf{m}_1(x)$ and $\mathbf{m}_2(x)$. To measure distance we will use a Riemannian discretization of the $L^2$ norm
\begin{equation}
    \| \mathbf{m}_1(x,t) - \mathbf{m}_2(x,t)\| = \sqrt{\int_{-L/2}^{L/2} | \mathbf{m}_1(x,t)-\mathbf{m}_2(x,t)|^2 dx} \approx \sqrt{ a \sum_{i=1}^N | \mathbf{m}_1(x_i,t)-\mathbf{m}_2(x_i,t)|^2},
\end{equation}
\noindent
where $a$ is the lattice spacing. Note that this distance is upper-bounded due to the unit-length constraint $|\mathbf{m}|^2=1$. At each point, the deviation is largest when the two fields are antiparallel at each point, e.g. $|(1,0,0) - (-1,0,0)|^2 = 4 \quad \forall x_i$. The upper bound of the norm of two configurations is
\begin{equation}
     \sqrt{ a \sum_{i=1}^N | \mathbf{m}_1(x_i)-\mathbf{m}_2(x_i)|^2}=\sqrt{ a 4N }=2\sqrt{L}.
\end{equation}

\textbf{Reference sets.} We estimate $\lambda_{FT}$ from a set of $M+1$ trajectories $\{\mathbf{m}_i(x,t)\}_{i=1}^{M+1}$, where $\mathbf{m}_1$ will be the reference trajectory. It is necessary to note that the estimate depends on the set chosen. The sets we will consider are
\begin{itemize}
    \item the thermal set. In this case the reference trajectory is generated via the Metropolis algorithm as described above. This has alerady been discussed in the main text.
    \item the smooth set. In this case the reference trajectory genereted from numerically evolving an analytic bump initial condition
\begin{equation}
    \mathbf{m}=\big( \mathrm{sin(\theta)}\mathrm{cos(\phi)}, \mathrm{sin(\theta)}\mathrm{sin(\phi)}, \mathrm{cos(\theta)}\big), \quad \phi = 2\pi x/L, \quad \theta =\pi e^{-x^2}.
\end{equation}
\noindent
\end{itemize}
In either case, the perturbations of $\mathbf{m}_1$ are performed in the same way. We perform a position-dependent rotation $\mathbf{m}_j(x) = R(x) \mathbf{m}_1(x)$, where $R(x) = \mathrm{exp}(\varepsilon A(x))$, and $A(x)$ is a random skew-symmetric matrix with a small parameter $\varepsilon \propto \delta_0(\mathbf{m}_1,\mathbf{m}_j)$. A general form for $A(x)$ can be written in Fourier space 
\begin{equation}
    A(x) = \sum_{k=1}^K \mu_k \, \mathrm{cos}\left(\frac{2\pi x}{L} + \varphi_k\right),
\end{equation}
\noindent
where $\mu_k \sim \mathcal{N}(0,\sigma_{\mu_k})$ are normally distributed, while $\varphi_k$  are uniformly distributed on $[0,2\pi)$. This is a convenient representation of $A(x)$ as it manifestly preserves the unit norm after application $|\mathbf{m}_j(x)\cdot \mathbf{m}_j(x)|=1$. We use $K=12$ and $\sigma_{\mu_k}=1/k^{2\beta}$, however $\varepsilon$ is still free to fix the average $\delta_0(\mathbf{m}_1,\mathbf{m}_j)$. $R(x)$ can then be determined numerically by computing the exponent of $A(x)$.

\begin{figure}[h!]
    \centerline{
    \includegraphics[width=1.0\linewidth]{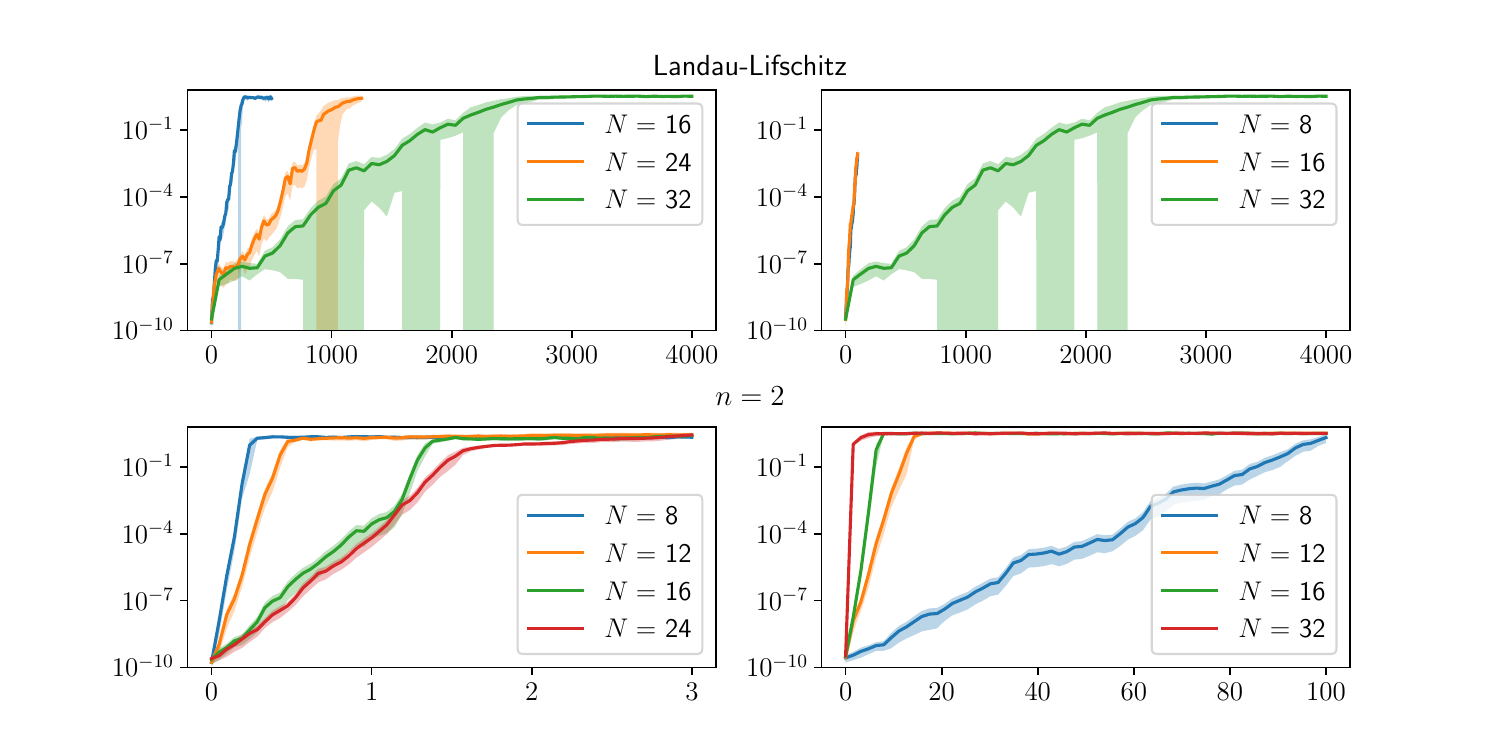}
    }
    \caption{FTLE estimation for discretized models with a smooth initial configuration with fixed lattice spacing $a=5/16$ (left panels) and with fixed $L=10$ (right panels). The top row shows the LL model, while the bottom shows $n=2$. Shaded regions show $\pm1\sigma$ deviations from 30 perturbed trajectories. }
\label{fig:FTLE_LL}
\end{figure}

For smooth configurations we checked two different limiting procedures. In the first procedure we fix the lattice spacing $a=5/16$ and pick $L=Na$, corresponding to $L\to\infty$ when $N\to\infty$. The second procedure is the continuum limit, where we fix $L=10$ and the lattice spacing $a = L/N$ is decreased. As can be seen in Fig.\,\ref{fig:FTLE_LL}, in the case of LL both procedures give a Lyapunov exponent which is diminished with $N$, suggesting regular behavior.

The case of $n=2$ is less clear. The fixed lattice spacing $a$ limiting procedure yields a $\lambda_{FT}$ that depends on the lattice spacing chosen $a$. On the other hand, fixing $L$ and decreasing the lattice spacing does not show a convergent Lyapunov exponent, as would be expected for a chaotic model. This signals that there is a strong energy transfer from the low frequency sector to the high frequency sector up until the maximal frequency $\pi N/L$ is hit. An increase of $N$ allows for the energy to be transferred to higher frequencies, which increases the exponential deviation of trajectories. Although this kind of dynamics signals irregular behavior, we cannot make a definite conclusion on the Lyapunov exponent. An accurate computation of $\lambda_{FT}$ would have to include alterations of the numerical scheme that could handle the strong energy transfer appropriately, however, this is beyond the scope of this work.

\newpage

\newpage
\section{Evaluation of momentum} \label{appendix:momentum}
In this section, we discuss various approaches to compute the value of the momentum charge numerically. The issue lies in the fact that we have performed a discretization of a continuum model, for which the momentum is well defined. However, if the discretized fields are not sufficiently smooth, the evaluation is inaccurate. We will present three different ways of computation.

The first method is direct integration of the charge density
\begin{equation} \label{eq:momentum_1}
    P=\int_{-L/2}^{L/2} \mathrm{d}x \frac{\mathbf{m}\cdot(\mathbf{m}_{x}\times\mathbf{m}_{xx})}{ (\mathbf{m}_x \cdot \mathbf{m}_x)^2}.
\end{equation}
\noindent
To evaluate this expression we use a spectral approach to compute the derivatives in Fourier space by applying the operator $\partial_x^n = (\mathrm{i}k)^n$. This approach leads to exponential convergence in the $N$ for analytic functions \cite{trefethen2000spectral}, which can be satisfied for special (approximately smooth) initial conditions or at large $\beta$. However, for noisy fields (i.e. low $\beta$) the second derivative is the leading contribution of the error.

\noindent
\textbf{Geometric picture.} The second approach is a geometric one. Since the total momentum is proportional to the area enclosed by the solid angle \cite{haldane1986geometrical} we can try to directly integrate the area enclosed by the curve on the sphere. We map the field to a curve on the unit sphere $\mathcal{S}^2$
\begin{equation}
    \mathbf{m}(x) = ( \mathrm{sin}\theta(x) \mathrm{cos}\varphi(x), \mathrm{sin}\theta(x) \mathrm{sin}\varphi(x), \mathrm{cos}\theta(x)),
\end{equation}
\noindent
and directly integrate the area enclosed by the curve
\begin{equation}
    A = \int \varphi_x(x)[1-\cos\theta(x)] \mathrm{d}x \propto P.
\end{equation}
\noindent
We find this works better than direct evaluation of the triple product because one needs to differentiate $\varphi$ only once. However, the mapping to spherical coordinates introduces multiple issues. The first is that $\varphi$ is defined up to modulo $2\pi$ which introduces the need to "unwrap" it in order to compute the derivative $\varphi_x(x)$ without spurious jumps of $2\pi/a$. Connected with this is the second issue -- because the frequency of sampled points is $1/a$ this gives a bound on how accurately we can unwrap $\varphi(x)$. For example, when a jump over $\pm \pi$ is observed one cannot definitely conclude if the field crossed the threshold if the derivative is correct. Thus, assumptions on the noise level along the curve must be made. Finally, the trajectory can have a non-zero winding number on the unit sphere, which must be accounted for, or a large derivative can appear at $\varphi_x(\pm L/2)$. This method outperforms the direct evaluation of the triple product if all the issues are correctly taken into account.

\noindent
\textbf{Pole formula.} The third approach is motivated by \cite{haldane1986geometrical}. We begin by initializing the pole $\hat{z}=(0,0,1)$ (which can be chosen arbitrarily). Then the momentum can be expressed as
\begin{equation}
    P = \int_{-L/2}^{L/2} \frac{ (\mathbf{m}_x \cdot \mathbf{m}) \cdot \hat{z}}{1 + \mathbf{m} \cdot \hat{z}} \mathrm{d}x
\end{equation}

\noindent
This formulation of the momentum elucidates an additional problem with its actual computation -- the divergence when the field anti-aligns with the pole $\mathbf{m}\cdot\hat{z} = -1$. For non-smooth configurations, one can expect for a point to be arbitrarily close to $\hat{z}$ which leads to a divergence. However, in practice, for smooth initial configurations this method seems to outperform the other two, so we will discuss it here.

\begin{figure}[h!]
    \centerline{
    \includegraphics[width=0.5\linewidth]{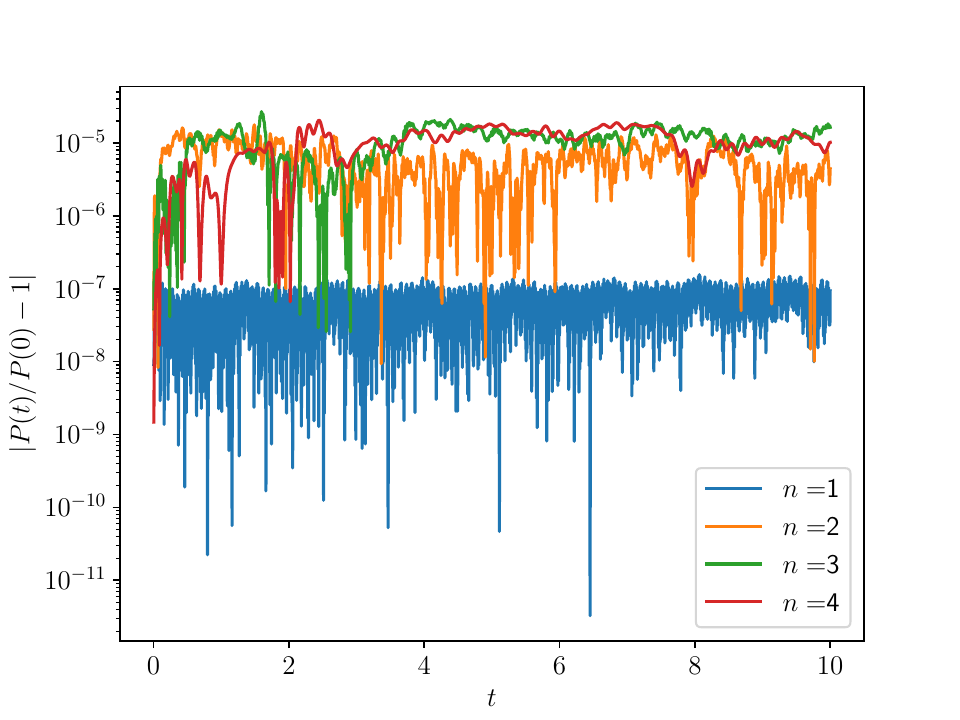}
    \includegraphics[width=0.5\linewidth]{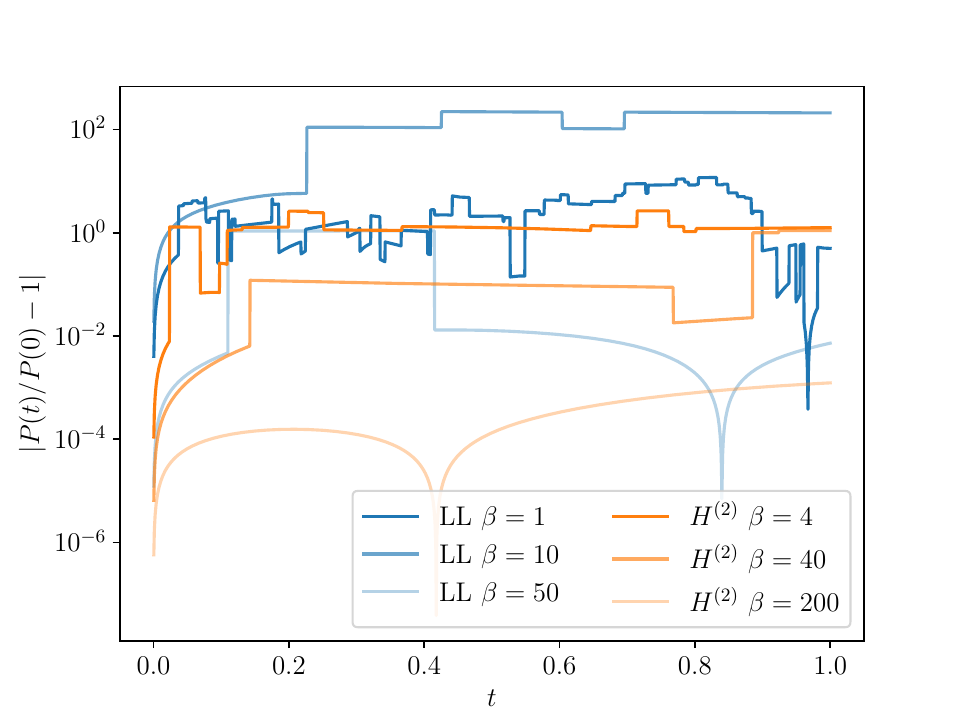}
    }
    \caption{\textbf{Left.} Relative conservation of momentum with a smooth initial configuration for increasing $n$. Computed using the pole formula. ($N=256,\,L=10$, integrator tolerance $10^{-8}$). \textbf{Right.} Relative conservation of momentum with a thermal initial configuration with increasing $\beta$. Calculated using the geometric picture.  ($N=256$, $L=N$, integrator tolerance $10^{-8}$).}
\label{fig:conservation_torsion_relative}
\end{figure}

We check the conservation of momentum by evolving a smooth initial state with non-zero momentum $P \propto 2\pi(1-\mathrm{cos}\theta)$
\begin{equation}
    \mathbf{m}=( \mathrm{sin}\theta \mathrm{cos}\phi, \mathrm{sin}\theta \mathrm{sin}\phi, \mathrm{cos}\theta), \quad \phi=2\pi x/L.
\end{equation}
for various $n$. An example is given on the left panel of Fig.~\ref{fig:conservation_torsion_relative}, which shows that the torsion is indeed conserved well for smooth initial conditions. We also perform the same test for thermal states which reveals that for a choice of intermediate $\beta$ the momentum is comparatively preserved much worse (right panel on Fig.~\ref{fig:conservation_torsion_relative}).
\noindent

\end{document}